\documentstyle[preprint,tighten,prc,eqsecnum,aps]{revtex}

\begin{document}


\preprint{IU/NTC 94-02 //OSU--94-113}

\title{TWO-LOOP CALCULATIONS WITH VERTEX CORRECTIONS IN THE WALECKA
MODEL}
\author{Brian D. Serot}
\address{Physics Department and Nuclear Theory Center\\
         Indiana University,\ \ Bloomington,~Indiana\ \ 47405}
\author{Hua-Bin Tang}
\address{Department of Physics\\
         The Ohio State University,\ \ Columbus,~Ohio\ \ 43210}

\date{April, 1994}
\maketitle
%

\begin{abstract}
Two-loop corrections with scalar and vector form
factors are calculated for nuclear
matter in the Walecka model.
The on-shell form factors are derived from vertex
corrections within the framework of the model and are highly damped at
large
spacelike momenta.
The two-loop corrections are evaluated first by using the
one-loop parameters and mean fields and then  by refitting
the total energy/baryon to empirical nuclear matter saturation
properties.
The modified two-loop corrections are significantly smaller than those
computed with bare vertices.
Contributions from the anomalous isoscalar form factor of the nucleon
are
included for the first time.
The effects of the implicit density dependence of the form
factors, which arise from the shift in the baryon mass, are also
considered.
Finally, necessary extensions of these calculations are discussed.

( Figures are available by E-mail to HTANG@MPS.OHIO-STATE.EDU )
\end{abstract}


\narrowtext

\section{INTRODUCTION}

The traditional theory of
nuclear structure is based on the Schr\"{o}dinger equation
with a nucleon--nucleon (NN) potential that has its origin in meson exchange.
In this nonrelativistic approach, one fits the NN potential to the empirical
properties of the deuteron and to low-energy NN scattering data, and one then
attempts to predict the behavior of many-nucleon systems.
A natural and
appealing generalization of this approach is to incorporate
special relativity by using
a relativistic quantum field theory with explicit meson
and baryon degrees of freedom.
These degrees of freedom are chosen because
they are the most efficient for describing low- and medium-energy nuclear
experiments.
This generalization allows us to study interacting,
relativistic, nuclear many-body systems,
which future experiments will examine.
The two basic questions are: what
kind of field theories should we use, and how well can we describe
nuclear systems using field theoretical
models with hadrons, which are actually particles with internal structure.
These are broad and hard questions that
can be answered only after intensive and systematic investigation.

Renormalizable relativistic quantum field theories with hadronic degrees
 of freedom, often called {\em quantum hadrodynamics\/} or QHD,
have been studied for some time \cite{WALECKA74,SEROT86,SEROT92}.
At the level of the mean-field theory (MFT) and one-loop approximation
(``relativistic Hartree approximation'' or RHA), these models can
reproduce nuclear matter saturation and can realistically describe
many bulk and single-particle properties of finite
nuclei \cite{SEROT86,SEROT92,FURNSTAHL87}.
The dynamical assumption behind renormalizability in QHD is that the
quantum vacuum and the internal structure of the hadrons can be described in
terms of hadronic degrees of freedom alone.
This assumption must ultimately break down at very short distances,
and its limitations can be
tested by explicit calculations.

Some approximate QHD calculations beyond the one-loop
level\cite{FURNSTAHL89,WEHRBERGER90,LIM90,TANAKA92}
indicate large vacuum corrections, and even the validity of
the one-loop vacuum contribution has been questioned \cite{COHEN89}.
However, it is unlikely that we can calculate consistently
beyond the mean-field level by
introducing {\em ad hoc\/} procedures, such as
including only the nucleons in the Fermi sea and simply throwing away
Dirac-sea contributions.
Indeed, it is already known
that vacuum contributions are indispensable for maintaining the conservation
of momentum and the electromagnetic current at the level of the
random-phase approximation
(RPA)\cite{FURNSTAHL89b,DAWSON90}.
These results imply
that we must develop practical and reliable
techniques that go beyond the MFT and that include vacuum dynamics.

A straightforward two-loop approximation for nuclear
matter in the Walecka model was examined in Ref.~\cite{FURNSTAHL89},
where large vacuum corrections were found.
The loop expansion does not appear to be convergent or asymptotic
in any sense.
This is not a surprise if we notice
that the two-loop corrections are essentially perturbative in the large
couplings.
An alternative expansion in
terms of meson loops (after integrating out the baryon fields) encounters
the well-known ghost problems\cite{WEHRBERGER90,LIM90,TANAKA92}.
Even after the ghost poles are removed by Redmond's ``surgical''
procedure\cite{REDMOND58},
putting aside for the moment its
physical significance, the vacuum contributions are
still big\cite{TANAKA92,TANG93}.
These calculations suggest that vertex corrections (and short-range
correlations) should
be included to compute vacuum loops reliably in QHD
theories\cite{MILANA91}.

Since hadrons have internal substructure,
one could argue for the use of nonlocal couplings and the introduction of
{\em ad hoc\/} vertex form factors.
This procedure was adopted in Refs.~\cite{PRAKASH91} and
\cite{FRIEDRICH92}.
We remark that the introduction of form factors
implies nonrenormalizability of the theory from the outset,
and thus the renormalization
procedure in Refs.~\cite{PRAKASH91} and \cite{FRIEDRICH92} needs
justification.
(In other words, various subtractions were made in the computation of the
energy that arise naturally only in a renormalizable theory.)
This {\em ad hoc\/} procedure
also says nothing about off-shell or density-dependent effects,
and poses difficulties when one tries to construct approximations that obey
the desired conservation laws.
Moreover, although
QCD in principle gives a complete description of nucleon structure,
some part of the internal properties of the nucleon, particularly at large
distances, must be equivalent to that provided by virtual hadron loops.
These are the physical effects we intend to study in this paper.

As pointed out recently by Milana\cite{MILANA91},
a theory with baryons and
vector mesons contains proper vertex functions that are highly damped at
large spacelike momentum transfers, due to the contributions from virtual
{\em bremsstrahlung} summed to all orders.
This damping arises from
the long-range (infrared) structure of the vertex and so should be calculable
within  the QHD framework.
Since the QHD theory is renormalizable, the vertex
function can be
expressed in terms of the couplings and masses of the theory, with
no {\em ad hoc\/} parameters.
Although it may be impossible to achieve a truly
quantitative description of the hadron structure within a QHD model,
due to the complicated nature of the vertex functions,
our goal is to understand at least
the qualitative features of the vertex functions implied in a hadronic theory
and to see how these features affect calculations with vacuum loops.

In a recent calculation\cite{ALLENDES92},
the on-shell vector form factor was studied in
a model with baryons and vector mesons.
The behavior at large spacelike momentum transfer $Q$ was determined by
the leading logarithmic infrared behavior, which arises from the sum of all
diagrams with vector ladders and crossed ladders across the hard vertex.
At small $Q$, the lowest-order vertex correction was used, and the complete
vertex function was constructed by
interpolating smoothly between the low- and high-momentum transfer regimes.
(See Ref.~\cite{KREIN93} for a similar analysis with a somewhat different
strategy.)

In this paper we extend the method of Ref.~\cite{ALLENDES92}
to the scalar--baryon
vertex and apply the form factors so obtained to two-loop calculations
in the Walecka model, with the lagrangian density
\begin{eqnarray}
{\cal {L}} & = & {\overline\psi} (i{\rlap/ \partial}
      - g_{\rm v} \rlap/{V}-M+g_{\rm s}\phi)\psi  \nonumber \\
        & &      {}+\case{1}{2}(\partial_{\mu} \phi \partial^{\mu} \phi
         - m_{\rm s}^2 \phi^2)          \nonumber \\
            &   &  {}- \case{1}{4}F_{\mu\nu}F_{\mu\nu}
                  + \case{1}{2}m _{\rm v}^2 V_{\mu}V^{\mu}
                    + \delta {\cal {L}} \ ,
\end{eqnarray}
where $F_{\mu\nu}=\partial^{\mu} V^{\nu}-\partial^{\nu} V^{\mu}$ and
$\delta \cal {L} $ contains the counterterms.
We observe that the {\em off-shell\/} vertex functions should be used
in a fully satisfactory calculation with loops.
A full off-shell calculation is quite complicated, however,
as one needs to know the off-shell behavior of the vertices at all spacelike
momenta, as well as the modification of the vertices in the presence of
valence nucleons at finite density.
Calculations exploring these off-shell vertex functions are in
progress\cite{ALLENDES93}.
Here, as a first step, we use an on-shell approximation,
in which the off-shell vertex functions are
replaced by their on-shell forms at zero density.
This procedure is analogous to that used in Refs.~\cite{PRAKASH91} and
\cite{FRIEDRICH92}, where parametrized, on-shell form factors were used
at the vertices, except that we use form factors obtained from within our
model.
Note also that the form factors used in Refs.~\cite{PRAKASH91} and
\cite{FRIEDRICH92}
were chosen to have simple momentum dependence, which allowed the two-loop
energy to be calculated directly from the results of
Ref.~\cite{FURNSTAHL89}; in contrast, we allow for an arbitrary momentum
dependence in the form factors and implement a {\em totally new\/}
renormalization procedure for the two-loop contributions.
Moreover, the anomalous vector form factor is included here
in the two-loop results for the first time.
Finally, we also estimate the effects of the implicit density
dependence due to the effective nucleon mass $M^{\scriptstyle \ast}$
that appears in the integrals that define the form factors.

To include the form factors
systematically, we apply the method of Freedman and
McLerran\cite{FREEDMAN77} to nuclear matter in the Walecka model.
One of the approximations that can be developed from this formalism is a
two-loop approximation with one dressed vertex\cite{TANG93}.
Using this approximation,
we calculate the corrections to the one-loop energy: first by using the
coupling parameters and mean fields determined at the one-loop level
and then by refitting
the total energy to the empirical nuclear matter saturation properties.
We find that the behavior of
the two-loop corrections is significantly improved
compared to that obtained with bare vertices\cite{FURNSTAHL89}.
(This agrees with results obtained in calculations with {\em ad hoc\/} vertex
functions\cite{PRAKASH91,FRIEDRICH92}.)
The contribution from the vector anomalous
vertex is small compared to those from the vector
charge and scalar vertices, but not negligible.
The effects of the $M^{\scriptstyle\ast}$ dependence and of
the uncertainty in our interpolations
for the form factors are modest compared to the overall size of the two-loop
corrections, but are nevertheless significant on the scale of the nuclear
matter binding energy.
These new effects, which are usually neglected in studies using
{\em ad hoc\/} form factors, are therefore
important for a detailed description of the
saturation properties of nuclear matter, because of the sensitive
cancellations in the energy that occur near equilibrium density.
It will certainly be necessary to extend our calculations to include the
integrations over the off-shell vertex functions before any definitive
statements can be made about nuclear matter saturation in this model.

Some remarks are in order about our calculation
of vacuum fluctuations involving the scalar field.
Here we treat the scalar field that represents
the $\sigma$ meson  as ``elementary'', although it should, as generally
believed, be considered as simulating the exchange of two correlated
s-wave pions\cite{DURSO80,LIN89,LIN90,FURNSTAHL93}.
Our goal in this work is to develop tools for calculating vacuum corrections
in a model with an NN interaction that has short-range repulsion and
mid-range attraction, as is empirically observed.
The Walecka model is a simple one that satisfies this constraint.
A more complete discussion of the role of chiral symmetry and the dynamical
generation of the mid-range NN attraction, together with the implications for
vacuum loops, should be considered as a necessary refinement of the work
presented here\cite{SEROT92,FURNSTAHL93}.
It is possible that a more detailed treatment could modify our
results significantly,
but it is also possible that a simple scalar field could remain an
adequate approximation to correlated two-pion exchange even
for the calculation of vacuum loops.
We leave these extensions as topics for future study.

The rest of this paper is organized as follows: In Sec.~\ref{sec:two}  we
discuss the renormalization of the energy corrections in the two-loop
approximation with one dressed vertex.
The finite expressions that are used in our numerical calculations are
generated.
In Sec.~\ref{sec:three} we present our
approximations for the vertex functions and the method for determining
the on-shell form factors.
The results for the energy of nuclear matter with the vertex functions fixed
at their free-space forms are given in Sec.~\ref{sec:four},
and we study the effects
of medium-modified vertices in Sec.~\ref{sec:five}.
Section VI is a summary, and some
technical details are included in an Appendix.

%

 \section{TWO-LOOP APPROXIMATION WITH VERTEX CORRECTIONS} \label{sec:two}

To include vertex functions in loop calculations of the nuclear matter
energy density, it is convenient to apply the method of
Freedman and McLerran\cite{FREEDMAN77},
in which the thermodynamic potential is
constructed as a function of the full, connected propagators and proper
vertices that satisfy
the Schwinger--Dyson equations.
By truncating the expansion of the thermodynamic
potential and the Schwinger--Dyson
equations appropriately, we can obtain various well-known
approximations, such as the MFT,
the RHA, the straightforward two-loop approximation\cite{FURNSTAHL89},
the relativistic RPA, {\em etc}.
Details of these formal procedures are reported
elsewhere\cite{TANG93,SEROT94}.
Here we merely quote the results for the two-loop
approximation with one dressed vertex.

Although the method of Freedman and McLerran provides for systematic
truncation, it says nothing about the best
way to approximate the exact thermodynamic potential in the case of strong
couplings.
It is also possible to truncate the expansion to include dressed vertices at
{\em both\/} ends of the two-loop diagram.
Here we shall take a conservative approach and include only one dressed
vertex; if this gives adequate suppression of the vacuum loops, two dressed
vertices will give even more.
The question of which of these truncations is a better starting point can
be answered only by including the next term in the expansion,
and we leave this as a topic for future study.

At the one-loop
level (or RHA), the nuclear matter energy density can be written
as\cite{SEROT86}
\begin{eqnarray}
     {\cal E}^{(1)} &=&
     {g _{\rm v}^2\over 2m _{\rm v}^2}\,
        \rho_{\scriptscriptstyle\rm B}^2
       + {m_{\rm s}^2\over 2g_{\rm s}^2}\,
       (M -M^{\scriptstyle\ast})^2      \nonumber \\
       & &
     {}+ {\gamma\over (2\pi)^3}\! \int^{k_{\scriptscriptstyle\rm F}}
       {\kern-.1em}{\rm d}^3{\kern-.1em}{k} \, E^*_{\rm k}
          + \Delta {\cal E} (M^{\scriptstyle\ast}) \, ,
          \label{eq:endens}
\end{eqnarray}
where $E^*_{\rm k} \equiv ({\bf k}^2 + {M^{\scriptstyle\ast}}^2)^{1/2}$
and $M^{\scriptstyle\ast}=M-g_{\rm s} \phi_{\scriptscriptstyle 0}$, with
$\phi_{\scriptscriptstyle 0}$ the average scalar field.
The spin-isospin degeneracy $\gamma = 4$
for nuclear matter and $\gamma = 2$ for
neutron matter, and the mean vector field has been eliminated using
 \begin{equation}
 V_{\scriptscriptstyle 0} = {g_{\rm v} \over m _{\rm v}^2}\,
    \rho_{\scriptscriptstyle\rm B} \equiv {g_{\rm v} \over m _{\rm v}^2}\,
          {\gamma k_{\scriptscriptstyle\rm F}^3\over 6 \pi^2} \, ,
 \end{equation}
 since it is a constant of the motion.
 The one-loop vacuum correction (``zero-point energy'') is
\widetext
 \begin{eqnarray}
     \Delta {\cal E} (M^{\scriptstyle\ast}) &=&
       {}-{1\over 4\pi^2} \left\{ M^{\scriptstyle\ast}{}^4
            \ln (M^{\scriptstyle\ast}/M) + M^3
            (M - M^{\scriptstyle\ast}) -
         \case{7}{2} \,M^2 (M - M^{\scriptstyle\ast})^2 \right.
               \nonumber \\[2pt]
       & &\qquad\qquad\qquad \left.
            {}+ \case{13}{3}\,
            M (M - M^{\scriptstyle\ast})^3
          - \case{25}{12} \,(M- M^{\scriptstyle\ast})^4 \right\}
         \, ,
               \label{eq:oldvf}
 \end{eqnarray}
\narrowtext
\noindent
where the renormalization conditions of Chin\cite{CHIN77} have been used,
and we have assumed that $\gamma = 4$ for the Dirac sea.

 The two-loop approximation with one dressed vertex obtained using
 the method of Freedman and McLerran is most transparently described by
 Feynman diagrams.
 The two-loop corrections to the RHA
 energy density can be drawn as in Fig.~\ref{FM2loop}, where
 $\Sigma$ denotes the
 renormalized proper baryon self-energy shown in Fig.~\ref{fig:sigma},
 and $\Lambda_{\rm s}$ ($\Lambda_{\rm v}$) stands for the dressed scalar
 (vector) vertex.
 The square brackets indicate that the enclosed subdiagrams are renormalized
 by the inclusion of the appropriate subtractions.
 (We follow the diagrammatic conventions of Ref.~\cite{FREEDMAN77}.)

 The propagators in Fig.~\ref{FM2loop} are as follows:
 The baryon Hartree propagator is
 \begin{equation}
 G^*(k) = G_{\scriptscriptstyle \rm F} ^*(k) +
          G_{\scriptscriptstyle\rm D} ^*(k) \, ,
 \end{equation}
 where
 \begin{eqnarray}
 G_{\scriptscriptstyle \rm F} ^*(k)  &= &\frac{\rlap/{\mkern-1mu k}
        +M^{\scriptstyle\ast}}{k^2
        - {M^{\scriptstyle\ast}}^2 +i\epsilon} \, , \\
 G_{\scriptscriptstyle\rm D} ^*(k)  &= &
       \frac{i\pi}{E^*_{\rm k}}(\rlap/{\mkern-1mu k}
       +M^{\scriptstyle\ast}) \delta (k^0 -E^*_{\rm k})
          \theta (k_{\scriptscriptstyle\rm F}-|{\bf k}|) \, ,
 \end{eqnarray}
 and the momentum has been shifted to eliminate
$V_{\scriptscriptstyle 0}$, which is permissible
 for the evaluation of closed loops\cite{FURNSTAHL89}.
 The free (noninteracting) scalar and vector propagators are
 \begin{eqnarray}
 \Delta ^0(q)     &= & {1\over q^2-m_{\rm s}^2+i\epsilon} \ , \\
  D^0_{\mu\nu}(q) &= & \left ( -g_{\mu\nu} +{q_{\mu}q_{\nu}\over m _{\rm v}^2}
                         \right ) D^0(q)  \ , \\
  D^0(q)          &= & { 1\over q^2-m _{\rm v}^2+i\epsilon }\ .
 \end{eqnarray}

 By separating the renormalized subdiagrams
 into unrenormalized parts and counterterm contributions (CTC), we
 obtain the more familiar diagrams in Fig.~\ref{2lp_fig}, where the CTC
 are not shown, since they can be determined easily from Fig.~\ref{FM2loop}.
 Evidently, an exact evaluation of these loops requires the knowledge of the
 off-shell proper vertices $\Lambda_{\rm s}$
 and $\Lambda _{\rm v}^\mu$, including their
 explicit density dependence.
 These off-shell vertices are currently under investigation\cite{ALLENDES93},
 but in this work, as a first
 approximation,
 we will replace them by their on-shell forms at zero density:
 \begin{eqnarray}
   ig_{\rm s} \Lambda_{\rm s} (q) & =&
          ig_{\rm s} F_{\rm s} (-q^2) \ ,\label{svtx} \\
   -ig_{\rm v} \Lambda _{\rm v}^{\mu}(q) & =&
         -ig_{\rm v} \left [F_{{\rm v}1}  (-q^2) \gamma^{\mu}
         \right. \nonumber \\
         & &\qquad\qquad {}+\left.
           iF_{{\rm v}2}  (-q^2)\sigma^{\mu\nu} q_{\nu}\right ] \ ,
         \label{vvtx}
 \end{eqnarray}
 where $\sigma^{\mu\nu}=i[\gamma^{\mu},\gamma^{\nu}]/2$,
and $q^\mu$ is defined as
 the {\em incoming\/} momentum transfer at the vertex.
 Note that in the calculation of the energy, the on-shell form factors are
 needed only at {\em spacelike\/} momentum transfers,
 so we can use the results of Ref.~\cite{ALLENDES92}.
 We emphasize that a similar on-shell approximation (with {\em ad hoc\/} form
 factors) is commonly
 used in essentially all calculations of nuclear matter properties
 (see, for example, Refs.~\cite{PRAKASH91} and \cite{FRIEDRICH92});
 here, however, we will determine the on-shell momentum
 dependence within the context of our model.
 This also allows us to discuss some of the implicit density-dependent effects
 (contained in $M^{\scriptstyle\ast}$), which we examine in Sec.~V.

 With these considerations, we can translate the Feynman diagrams in
 Figs.~\ref{FM2loop}, \ref{fig:sigma}, and \ref{2lp_fig}
 to obtain the renormalized two-loop correction
\widetext
 \begin{eqnarray}
 {\cal E}^{(2)} & =&\frac{1}{2}g_{\rm s}^2
          \int\!{{\rm d}^4 k \over (2\pi)^4}\,
        {{\rm d}^4 q \over (2\pi)^4}\,{\rm tr}
           \left[ G^*(k)G^*(k+q)\right] F_{\rm s}(-q^2)
         \Delta ^0(q) \nonumber \\
      & &{}- \frac{1}{2}g _{\rm v}^2 \int\!{{\rm d}^4 k \over
        (2\pi)^4}\,{{\rm d}^4 q \over (2\pi)^4}\,{\rm tr}
           \left[ G^*(k)\gamma_{\mu}
          G^*(k+q)\Lambda _{\rm v}^{\mu}(q)
            \right] D^0(q)  \nonumber \\
      & &{}+ i \int\!{{\rm d}^4 k \over (2\pi)^4}\,{\rm tr}
           \left[\Sigma_{\rm ct}
          G^*(k)\right]-\sum_{n=1}^{4}\alpha_n\phi_{\scriptscriptstyle 0}^n
                  -{\rm VEV} \ ,   \label{II9}
 \end{eqnarray}
\narrowtext
\noindent
where ``${\rm tr}$'' denotes a trace over spin and isospin indices, and
$\Sigma_{\rm ct}$ contains the counterterms for the baryon
self-energy, as given in the Appendix.
 In Eq.~(\ref{II9}), we have discarded
 the longitudinal $q_{\mu}q_{\nu}$ term in the vector
 propagator, since its contribution vanishes by baryon current conservation,
 as can be easily verified in our approximation.
 The vacuum-expectation-value (VEV) subtraction is obtained by replacing
 all of
 the Hartree propagators $G^*(k)$ with noninteracting Feynman propagators
 $G_{\scriptscriptstyle \rm F}^0 (k) =
 (\rlap/{\mkern-1mu k} - M + i\epsilon)^{-1}$
 and by omitting the quartic polynomial in
 $\phi_{\scriptscriptstyle 0}$.
 Note that some of the $\alpha_n$ counterterms contain both
 one-loop $[O(\hbar)]$ and two-loop $[O(\hbar^2)]$ contributions. The
 one-loop contribution comes from the renormalized  meson polarizations
 in Fig.~\ref{FM2loop}. See
 Eq.~(2.79) of Ref.~\cite{FURNSTAHL89} for the details of this separation.

 Due to our use of on-shell vertex functions in the two-loop integrals, the
 renormalization procedure is nonstandard and requires some discussion.
 In principle, the only unambiguous way to renormalize with vertex insertions
 is to use the off-shell vertices inside the integrals and include all
 required counterterms;
 this is clearly exceedingly difficult and motivates our
 simplified calculation using the on-shell vertices.
 We now observe that most of the
 effort in renormalizing the two-loop integrals
 with {\em bare\/} vertices involves the baryon propagators
 (in the self-energy and polarization loops).
 The noninteracting meson propagators
 simply follow along in the analysis, and although they determine which
 integrals diverge, they are otherwise innocuous.
 This suggests the following renormalization procedure with dressed vertices:
 \begin{itemize}%
\begin{enumerate}
 \item  We first carry out the renormalization of the vertex functions to
 arrive at finite, on-shell functions of the momentum transfer.
 This procedure is discussed in the next section, where we specify the
 detailed form of our model form factors.
 \item  These form factors are then associated with the corresponding meson
 propagators (which carry the same momenta).
 The form factors then serve only to modify the momentum dependence of the
 meson propagators, producing so-called
 ``M{\o}ller potentials''\cite{JAUCH76}.
 \item  The counterterms are then defined exactly as in the two-loop case with
 bare vertices\cite{FURNSTAHL89}, using the M{\o}ller potentials in place of
 the noninteracting boson propagators.
 \end{enumerate}
 \end{itemize}

 While this renormalization procedure may not be unique, it has the following
 advantages:
 \begin{itemize}%
\begin{enumerate}
 \item  Finite results are obtained with the same number of counterterms as in
 the original two-loop calculation.
 All of the counterterms are constants that are defined by vacuum amplitudes,
 and all counterterm subtractions are {\em local}.
 \item  All subintegrations for the meson polarizations and baryon
 self-energies
 satisfy the standard renormalization
 conditions\cite{FURNSTAHL89} for any choice of vertex functions.
 \item  The resulting energy density satisfies the same renormalization
 conditions as in the original two-loop
 calculation, for any choice of vertex functions.
 In particular, the original results are obtained automatically when the
 model vertices are replaced by bare vertices.
 Note that, even with the form factors included, nested and overlapping
 subtractions remain divergent, although the overall subtractions are finite,
 as are the baryon self-energy counter\-terms.
 \end{enumerate}
 \end{itemize}

 We remark that in a theory that postulates nonlocal vertices in the
 lagrangian,
 the definition of the renormalization counterterms is essentially arbitrary.
 In particular, since the overall subtractions to the energy are
 finite, there is no
 justification for stopping at $O(\phi_{\scriptscriptstyle 0}^4)$,
 and one can simply remove all
 vacuum fluctuations by fiat.
 (That is, one can include an infinite number of counterterm subtractions.)
 To our knowledge, the {\em only\/} justification for stopping at a quartic
 polynomial is that the underlying theory is renormalizable, so that terms of
 $O(\phi_{\scriptscriptstyle 0}^5)$ and higher can never be removed.

 Following Ref.~\cite{FURNSTAHL89}, we decompose the
 two-loop contribution into
 exchange, Lamb-shift, and vacuum fluctuation energies. Thus
 \begin{equation}
 {\cal E}^{(2)} = {\cal E}_{\scriptscriptstyle\rm EX}
   + {\cal E}_{\scriptscriptstyle\rm LS}
   + {\cal E}_{\scriptscriptstyle\rm VF} \ ,
 \end{equation}
 where
\widetext
 \begin{eqnarray}
  {\cal E}_{\scriptscriptstyle\rm EX} &= &
       \frac{1}{2}g_{\rm s}^2 \int\!{{\rm d}^4 k \over (2\pi)^4}\,
        {{\rm d}^4 q \over (2\pi)^4}\,{\rm tr}
         \left[ G_{\scriptscriptstyle\rm D} ^*(k)
         G_{\scriptscriptstyle\rm D} ^*(k+q)\right]
          \Delta ^0(q) F_{\rm s}(-q^2) \nonumber \\
     & & {}-\frac{1}{2}g _{\rm v}^2 \int\!{{\rm d}^4 k \over (2\pi)^4}\,
         {{\rm d}^4 q \over (2\pi)^4}\,{\rm tr}
           \left[ G_{\scriptscriptstyle\rm D} ^*(k)
          \gamma_{\mu} G_{\scriptscriptstyle\rm D} ^*(k+q)
          \Lambda _{\rm v}^{\mu}(q)\right]
            D^0(q)  \ , \\[6pt]
  {\cal E}_{\scriptscriptstyle\rm LS} & =&
          g_{\rm s}^2\int\!{{\rm d}^{\tau}k\over (2\pi)^4}\,
        {{\rm d}^{\tau}q\over (2\pi)^4}\,{\rm tr}
         \left[ G_{\scriptscriptstyle \rm F} ^*(k)
        G_{\scriptscriptstyle\rm D} ^*(k+q)\right]
          \Delta ^0(q) F_{\rm s}(-q^2)
           + i \int\!{{\rm d}^4 k \over (2\pi)^4}\,{\rm tr}
    \left[\Sigma_{\rm ct} G_{\scriptscriptstyle\rm D}^*(k)\right]
                   \nonumber \\
     & & {}- g _{\rm v}^2 \int\!{{\rm d}^{\tau}k\over (2\pi)^4}\,
             {{\rm d}^{\tau}q\over (2\pi)^4}\,{\rm tr}
           \left[ G_{\scriptscriptstyle \rm F} ^*(k) \gamma_{\mu}
      G_{\scriptscriptstyle\rm D} ^*(k+q)
          \Lambda _{\rm v}^{\mu}(q)\right]
            D^0(q) \ , \\[6pt]
 {\cal E}_{\scriptscriptstyle\rm VF} & =&
     \frac{1}{2}g_{\rm s}^2\int\!{{\rm d}^{\tau}k\over (2\pi)^4}\,
       {{\rm d}^{\tau}q\over (2\pi)^4}\,{\rm tr}
         \left[ G_{\scriptscriptstyle \rm F} ^*(k)
           G_{\scriptscriptstyle \rm F} ^*(k+q)\right]
        \Delta ^0(q) F_{\rm s}(-q^2)
        + i \int\!{{\rm d}^{\tau}k\over (2\pi)^4}\,{\rm tr}
           \left[\Sigma_{\rm ct}
          G_{\scriptscriptstyle \rm F}^*(k)\right]
                                       \nonumber \\
     & & {}-\frac{1}{2} g _{\rm v}^2
           \int\!{{\rm d}^{\tau}k\over (2\pi)^4}
        \,{{\rm d}^{\tau}q\over (2\pi)^4}\,{\rm tr}
           \left[ G_{\scriptscriptstyle \rm F} ^*(k) \gamma_{\mu}
       G_{\scriptscriptstyle \rm F} ^*(k+q)
          \Lambda _{\rm v}^{\mu}(q) \right]
            D^0(q) -\sum_{n=1}^{4}\alpha _n
          \phi_{\scriptscriptstyle 0}^n-{\rm VEV}  \ .
          \nonumber \\
     & & \label{vfe}
 \end{eqnarray}
 Here the divergent integrals have been regularized by writing them in $\tau$
 dimensions, with the limit $\tau
 \rightarrow 4$ taken after the divergences have been removed.
 Note that these energy densities are implicitly
 functions of the baryon density $\rho_{\scriptscriptstyle\rm B}$
and the mass parameter $M^{\scriptstyle\ast}$, with
 the latter to be determined by minimization of the full energy density.

 The exchange contribution
${\cal E}_{\scriptscriptstyle\rm EX}$ is finite.
 It is straightforward to work out the
 traces and arrive at the following form that is feasible for numerical
 integration using Gaussian quadrature:
 \begin{eqnarray}
  {\cal E}_{\scriptscriptstyle\rm EX} & =&
        \frac{1}{16\pi^4}\int_{0}^{k_{\scriptscriptstyle\rm F}}\!\!{\rm d}k
           \int_{0}^{k_{\scriptscriptstyle\rm F}} \!\!{\rm d}q\,
           {k^2q^2 \over E^*_{\rm k} E^*_{\rm q}}\,
              \int_{-1}^{+1} \!\!{\rm d}x \left[g_{\rm s}^2 F_{\rm s}(P^2)
              \frac{P^2+4M^{\scriptstyle\ast}{}^2}{P^2+m_{\rm s}^2}
                                   \right.\nonumber \\[4pt]
        &&\ \ \ \ \quad
           \left. {}+ 2g _{\rm v}^2F_{{\rm v}1} (P^2)
          \frac{P^2-2M^{\scriptstyle\ast}{}^2}{P^2+m _{\rm v}^2}
             +{6g _{\rm v}^2M^{\scriptstyle\ast}F_{{\rm v}2}
               (P^2)P^2\over P^2+m _{\rm v}^2} \right] \ ,
 \end{eqnarray}
 where we have defined
$P^2\equiv 2(E^*_{\rm k} E^*_{\rm q}-M^{\scriptstyle\ast}{}^2-kqx)$,
 with the dummy variables $k\equiv {| {\bf k}|}$ and  $q\equiv
 | {\bf q}|$.
 (Here and henceforth we take the spin-isospin degeneracy $\gamma = 4$.)
 Note that since the momenta flowing through the meson propagators are
 spacelike, one does not encounter any poles.

 The Lamb-shift contribution ${\cal E}_{\scriptscriptstyle\rm LS}$ can
be written as
 \begin{eqnarray}
  {\cal E}_{\scriptscriptstyle\rm LS} &\equiv&
        -i\int\!{{\rm d}^4 k \over (2\pi)^4}\mathop{\rm tr}\nolimits
           \left[ G_{\scriptscriptstyle\rm D} ^*(k)
          \Sigma _{\scriptscriptstyle\rm F}(k)\right] \\
        & = & \left[\Sigma_{\scriptscriptstyle\rm FA}
          (-M^{\scriptstyle\ast}{}^2)+M^{\scriptstyle\ast}
            \Sigma_{\scriptscriptstyle\rm FB}
          (-M^{\scriptstyle\ast}{}^2)\right]\, 4
            \int\!{{\rm d}^3k\over (2\pi)^3}
         \frac{M^{\scriptstyle\ast}}{E^*_{\rm k}}\,\theta
            (k_{\scriptscriptstyle\rm F}-|{\bf k}|) \nonumber \\
        & = & \left[\Sigma_{\scriptscriptstyle\rm FA}
         (-M^{\scriptstyle\ast}{}^2)+M^{\scriptstyle\ast}
            \Sigma_{\scriptscriptstyle\rm FB}
          (-M^{\scriptstyle\ast}{}^2)\right]\,
            \rho_{\rm s} (M^{\scriptstyle\ast} ;
         \rho_{\scriptscriptstyle\rm B}) \ ,
 \end{eqnarray}
 where the renormalized Feynman part of the baryon self-energy is
 \begin{eqnarray}
 \Sigma_{\scriptscriptstyle\rm F} (k) &\equiv&
           \Sigma_{\scriptscriptstyle\rm FA} (-k^2)
      +\rlap/{\mkern-1mu k} \Sigma_{\scriptscriptstyle\rm FB}(-k^2)
                 \nonumber \\
             & =&i \int{{\rm d}^{\tau}q\over (2\pi)^4}
            \left[ g_{\rm s}^2G_{\scriptscriptstyle \rm F}^*(k-q)
             \Delta^0(q)F_{\rm s}(-q^2)
                           \right. \nonumber \\
             &&\quad \quad \ \ \ \ \ \ \ \ \ \left.
               {}-g _{\rm v}^2 \Lambda _{\rm v}^{\mu}(q)
                 G_{\scriptscriptstyle \rm F}^*(k-q)
               \gamma_{\mu} D^0 (q)\right]
                    - \Sigma_{\rm ct}\ , \label{eq:slf}
 \end{eqnarray}
 and the specification of the counterterm coefficients
 \begin{equation}
 \Sigma_{\rm ct} \equiv -\zeta_{\scriptscriptstyle\rm N}(\rlap/ k-M)
       +\gamma_{\rm s}\phi_{\scriptscriptstyle 0}+M_{\rm c}
 \end{equation}
 is discussed in the Appendix.
 The scalar density of baryons is denoted by $\rho_{\rm s}$.

 By following the procedures in Ref.~\cite{FURNSTAHL89}, the vacuum
 fluctuation energy ${\cal E}_{\scriptscriptstyle\rm VF}$
     can be evaluated by expanding the integrands in
 Eq.~(\ref{vfe}) in powers of $(M^{\scriptstyle\ast}-M)$ using
 the algebraic identity
 \begin{equation}
 G_{\scriptscriptstyle \rm F}^*(k)=
        \sum_{i=0}^{n}(M^{\scriptstyle\ast}-M)^i
        [G_{\scriptscriptstyle \rm F}^0(k)]^{i+1}
 +(M^{\scriptstyle\ast}-M)^{n+1}
           [G_{\scriptscriptstyle \rm F}^0(k)]^{n+1}
        G_{\scriptscriptstyle \rm F}^*(k) \ .  \label{eq:Furry}
 \end{equation}
 Here $n$ can be any positive integer or zero,
 and $G_{\scriptscriptstyle \rm F}^0(k)$ is obtained from
 $G_{\scriptscriptstyle \rm F}^*(k)$ by replacing $M^{\scriptstyle\ast}$
 with $M$.
 The zeroth-order term in the
 expansion $[ {\cal E}_{\scriptscriptstyle\rm VF} (M)]$ is cancelled by the
 VEV, and the
 counterterms $\alpha_n$ are chosen as usual
 to cancel exactly the coefficients of the first {\em four\/}
 powers of $(M^{\scriptstyle\ast}-M)$,
 which minimizes many-body forces\cite{CHIN77,SEROT86}.
 The final result can be written as
 \begin{eqnarray}
  {\cal E}_{\scriptscriptstyle\rm VF}   & =&
      i(M-M^{\scriptstyle\ast})^5
         \int\!{{\rm d}^4 k \over (2\pi)^4}\,{\rm tr}
                  \left\{ [G_{\scriptscriptstyle \rm F} ^0(k)]^5
           G_{\scriptscriptstyle \rm F}^*(k)
           \Sigma^0_{\scriptscriptstyle \rm F}(k)
                  +[G_{\scriptscriptstyle \rm F} ^0(k)]^4
        G_{\scriptscriptstyle \rm F} ^*(k)\Lambda^0(k)\right\}  \nonumber  \\
          & & {}+ \frac{1}{2}(M-M^{\scriptstyle\ast})^5
            \int\!{{\rm d}^4 k \over (2\pi)^4}\,
        {{\rm d}^4 q \over (2\pi)^4}\,{\rm tr}
               \left\{ [G_{\scriptscriptstyle \rm F} ^0(k-q)]^3
          G_{\scriptscriptstyle \rm F} ^*(k-q) \right. \nonumber \\
         & &\qquad\qquad\qquad
         \times \left.\left( g _{\rm v}^2 \gamma_{\mu}
             [G_{\scriptscriptstyle \rm F} ^0(k)]^3\Lambda _{\rm v}^{\mu}(q)
               D^0(q)-g_{\rm s}^2 [G_{\scriptscriptstyle \rm F} ^0(k)]^3
             \Delta ^0(q) F_{\rm s}(-q^2)\right) \right \}  \nonumber \\
          & & {}+ \frac{1}{2}(M-M^{\scriptstyle\ast})^5
            \int\!{{\rm d}^4 k \over (2\pi)^4}\,
            {{\rm d}^4 q \over (2\pi)^4}\,{\rm tr}
               \left\{ [G_{\scriptscriptstyle \rm F} ^0(k-q)]^3
          G_{\scriptscriptstyle \rm F} ^*(k-q)\right. \nonumber \\
           & &\qquad\qquad\qquad\times \left. \left(g _{\rm v}^2
            \gamma_{\mu} [G_{\scriptscriptstyle \rm F} ^0(k)]^2
               G_{\scriptscriptstyle \rm F} ^*(k)
            \Lambda _{\rm v}^{\mu}(q)       D^0(q)
               -g_{\rm s}^2 [G_{\scriptscriptstyle \rm F} ^0(k)]^2
         G_{\scriptscriptstyle \rm F} ^*(k)\Delta ^0(q) F_{\rm s}(-q^2)
               \right) \right \} \ , \nonumber \\
          && \label{III6}
 \end{eqnarray}
 where $\Sigma_{\scriptscriptstyle\rm F}^0(k)$ is obtained from
 $\Sigma_{\scriptscriptstyle\rm F} (k)$
by replacing $M^{\scriptstyle\ast}$ with $M$,
 and the renormalized vacuum scalar vertex at zero momentum transfer,
 \begin{eqnarray}
 \Lambda^0(k) &\equiv&\Lambda^0_{\scriptscriptstyle\rm A} (-k^2)
   +\rlap/{\mkern-1mu k} \Lambda^0_{\scriptscriptstyle\rm B} (-k^2)
                    \nonumber \\
              & =&i \int{{\rm d}^{\tau}q\over (2\pi)^4}
       \left\{ g_{\rm s}^2[G_{\scriptscriptstyle \rm F}^0(k-q)]^2\Delta^0(q)
                       F_{\rm s}(-q^2) \right.   \nonumber \\
              &&\ \ \ \ \ \ \ \ \ \ \ \ \ \ \ \ \ \ \left.
       {}-g _{\rm v}^2 \Lambda _{\rm v}^{\mu}(q)
        [ G_{\scriptscriptstyle \rm F}^0(k-q)]^2\gamma_{\mu} D^0(q)\right\}
             +\frac{\gamma_{\rm s}}{g_{\rm s}} \ , \label{eq:vtx}
 \end{eqnarray}
 is discussed in the Appendix.
 Note that all integrals in Eq.~(\ref{III6}) are {\em finite}, even for point
 vertices.

 After working out the traces,
 one can perform a Wick rotation to Euclidean momenta and evaluate the
 angular integrals analytically, leading to
 ($\widetilde{{\cal E}}_{\scriptscriptstyle\rm VF} \equiv
  {\cal E}_{\scriptscriptstyle\rm VF} / M^4$)
 \begin{eqnarray}
 \widetilde{{\cal E}}_{\scriptscriptstyle\rm VF}
 & =&\frac{1}{2\pi^2}\int_{0}^{\infty}\!\! {\rm d}y\,
      \frac{(1-m^{\scriptstyle \ast})^5y}
           {(y+1)^5(y+{m^{\scriptstyle \ast}}^2)}
       \left\{\Sigma_{\scriptscriptstyle \rm {FA}}^0(y)
           \left[y^3-5(2+m^{\scriptstyle \ast})y^2
       +5(1+2m^{\scriptstyle \ast})y-m^{\scriptstyle \ast} \right]
           \right. \nonumber \\
       &&\qquad\qquad\qquad\quad\ \ \ \ \ \ \ \ \ \ \ \ \ \ \ \ \ \ \ \left.
            {}+\Sigma_{\scriptscriptstyle \rm {FB}}^0(y)
           \left[(5+m^{\scriptstyle \ast})y^2
        -10(1+m^{\scriptstyle\ast})y
      +1+5m^{\scriptstyle \ast}\right]y
           \right\} \nonumber \\
       &&{}+ \frac{1}{2\pi^2}\int_{0}^{\infty}\!\!{\rm d}y\,
        \frac{(1-m^{\scriptstyle \ast})^5y}
           {(y+1)^4(y+{m^{\scriptstyle \ast}}^2)}
          \left\{\Lambda^0_{\scriptscriptstyle\rm A}(y)
            \left[(4+m^{\scriptstyle \ast})y^2
       -2(2+3m^{\scriptstyle \ast})y+m^{\scriptstyle \ast}\right]
           \right. \nonumber \\
       &&\qquad\qquad\qquad\qquad\ \ \ \ \ \ \ \ \ \ \ \ \ \ \ \ \ \ \ \left.
            -\Lambda^0_{\scriptscriptstyle\rm B}(y)
            \left[y^2-2(3+2m^{\scriptstyle \ast})y
           +1+4m^{\scriptstyle \ast} \right]y\right\} \nonumber \\
       &&{}+\frac{1}{128\pi^4}\int_{0}^{\infty}\!\! {\rm d}s
            \int_{0}^{\infty}\!\! {\rm d}t\,
            \frac{(1-m^{\scriptstyle \ast})^5}
            {(t+1)^3(t+m^{\scriptstyle \ast}{}^2)}
           \left\{ \left[t^2-3(m^{\scriptstyle \ast}+1)t
               +m^{\scriptstyle \ast} \right]
               \left[u_1(s,t)A_1(s)\right. \right. \nonumber \\
        &&\qquad\qquad\qquad -
            \left. v_1(s,t)B(s)\right]+
            \left. \left[(m^{\scriptstyle \ast} +3)t^{\ }
          -3m^{\scriptstyle \ast} -1\right]
            \left[u_2(s,t)A_2(s) - v_2(s,t)B(s)\right]
                 \vphantom{\left[t^2]\!\right]}\right\}
                  \ . \nonumber \\
          && \label{vf:detail}
 \end{eqnarray}
 Here we have scaled all dimensional 
 variables with the nucleon mass, set
 $m^{\scriptstyle \ast} \equiv M^{\scriptstyle\ast}/M$, and defined
 \begin{eqnarray}
 A_1(s) & = &{g_{\rm s}^2 F_{\rm s}(s) \over s+m_{\rm s}^2}
       -{4g _{\rm v}^2F_{{\rm v}1} (s)\over s+m _{\rm v}^2}
             \ , \label{eq:a1} \\
 A_2(s)    & = &{g_{\rm s}^2 F_{\rm s}(s)\over s+m_{\rm s}^2}
      +{2g _{\rm v}^2 F_{{\rm v}1} (s)\over s+m _{\rm v}^2}
              \ ,  \label{eq:a2} \\
 B(s)    & = &{3g _{\rm v}^2 F_{{\rm v}2}(s)\over s+m _{\rm v}^2}
                  \ ,  \label{eq:bs}
 \end{eqnarray}
 and
 \begin{eqnarray}
 u_1(s,t) & = &{8st\over Z^{3}}-\frac{m^{\scriptstyle \ast} C(s,t)}
         {(1-m^{\scriptstyle \ast})^2}
                    -{5+m^{\scriptstyle \ast}\over Z}(s+t+1-Z) \ , \\[4pt]
 u_2(s,t) & = &{4st \over Z^3} (t-s-1)
                     +\frac{s-t+m^{\scriptstyle \ast}{}^2}
           {2(1-m^{\scriptstyle \ast})^2}C(s,t) \nonumber \\
          & &  {}+{1\over Z}\left[s-t+3-Z+{\textstyle {1\over 2}}
                (1+m^{\scriptstyle \ast})^2\right](s+t+1-Z)  \ ,  \\[4pt]
 v_1(s,t) & = & {4st\over Z^3}(t-s+1)-\frac{t-s+
          m^{\scriptstyle \ast}{}^2}
             {2(1-m^{\scriptstyle \ast})^2}C(s,t)
                        \nonumber \\
          & & {} -{1\over Z}\left[t-s+3-Z+{\textstyle {1\over 2}}
                    (1+m^{\scriptstyle \ast})^2\right](s+t+1-Z)  \ ,  \\[4pt]
 v_2(s,t) & = & {4st\over Z^3}(t+s+1)-\frac{m^{\scriptstyle \ast}
        (t+s+m^{\scriptstyle \ast}{}^2)}
                  {2(1-m^{\scriptstyle \ast})^2}C(s,t) \nonumber \\
          & & {} -{1\over 2Z}\left[(5+m^{\scriptstyle \ast})
            (s+t+m^{\scriptstyle \ast}{}^2-Z)
                  +9+2m^{\scriptstyle \ast}-3m^{\scriptstyle \ast}{}^2\right]
                  (s+t+1-Z)  \ , \nonumber \\
          & & \\[4pt]
 C(s,t)   & = & \sqrt{(s+t+m^{\scriptstyle \ast}{}^2)^2-4st}
                   -Z+{1-m^{\scriptstyle \ast}{}^2 \over Z}(s+t+1)  \ ,
 \end{eqnarray}
\narrowtext
\noindent
with $Z  \equiv \sqrt{(s+t+1)^2-4st} $.
We remark that the first two integrals in Eq.~(\ref{vf:detail}) correspond
precisely to the final two integrals in Eq.~(2.93) of
Ref.~\cite{FURNSTAHL89}.
The final integral in Eq.~(\ref{vf:detail}), however, was obtained here
with a new renormalization procedure, and thus it cannot be directly
compared to the quadrature in Ref.~\cite{FURNSTAHL89}.

 The integrals in Eq.~(\ref{vf:detail}) can be
 evaluated numerically using Gaussian quadrature.
 We found that splitting the integrations into two regions in each integral
 produced results that were accurate to better than 0.1\% with a moderate
 number of points ($\approx 32$ in each region).
 The largest uncertainty comes from the final integral in
 Eq.~(\ref{vf:detail}).
 We checked our computations by using two separate computer codes;
 all results for the nuclear matter energy presented below
 agreed to at least three digits.

\section{SCALAR AND VECTOR VERTEX FUNCTIONS} \label{sec:three}

In the preceding section, we
renormalized the two-loop energy with the approximation
that the fully off-shell vertices can be replaced
by their on-shell forms at zero
density, namely, Eqs.~(\ref{svtx}) and (\ref{vvtx}).
Here we evaluate these form
factors within the framework of our model.
Since an exact calculation of the vertex function is impossible at present,
we must make approximations.
Our strategy is to include in the on-shell vertex functions
the dominant physics that is accessible in a hadronic theory.

We begin with the well-known fact that the proper vertex function in QED falls
rapidly when the momentum $q^{\mu}$ entering on the photon
line becomes large\cite{SUDAKOV56,FISHBANE71}.
In particular, the asymptotic form for the on-shell vertex at large spacelike
momenta $q^2 < 0$ is
\begin{eqnarray}
{\overline u}(p_b) \Lambda^\mu u(p_a) &\longrightarrow&
    {\overline u}(p_b) \gamma^{\mu} u(p_a) \nonumber \\
   & & {}\times\exp \left[- \frac{e^2}{16\pi^2}
         \ln^2 \left(-{q^2\over m^2}\right)\right]
         \  , \label{eq:basic}
\end{eqnarray}
where $p_a^2 = p_b^2 = m_e^2$, and $m$ is an infrared regulator mass.
In a theory with a {\em massive\/} neutral
vector boson, the regulator mass $m$ is replaced by the boson mass, and
the electron charge $e$ becomes the vector coupling $g_{\rm v}$.
The physical origin of the strong damping is the large likelihood for
virtual {\em bremsstrahlung\/} of soft vector bosons.
In diagrammatic terms, the exponential arises from summing all ladders and
crossed ladders involving the exchange of soft bosons across the single hard
vertex, as shown in Fig.~\ref{fig:sv_diag}.
We emphasize that although the momentum transfer to the vertex
is large, the damping arises from the {\em infrared\/} structure of
the theory, as the required factors of $\ln^2(-q^2)$ are generated by
loop momenta that are on the order of the vector meson mass.
Thus it is reasonable to include this long-range vertex structure in a
renormalizable theory containing hadron loops.
Later work supports the assumption that non-leading logarithms
appear only as multiplicative
factors\cite{MUELLER79,COLLINS80,SEN81,STERMAN}.

It is easy to show that the exponential damping in Eq.~(\ref{eq:basic}) is
also reproduced by diagrams in which vector ladders and crossed ladders dress
a single (hard) scalar vertex, at least to leading logarithmic order.
Since the inclusion of higher-order vertex diagrams involving scalar
meson exchanges produces only ultraviolet (``hard'') logarithms,
as suggested by the work of Appelquist and
Primack\cite{APPELQUIST70}, these diagrams will not ruin the exponential
damping\cite{STERMAN}.

The general forms for the on-shell scalar and vector vertex functions at zero
density are given in Eqs.~(\ref{svtx}) and (\ref{vvtx}).
Following Ref.~\cite{ALLENDES92}, we assume
that the behavior of both vertices at small  $|q^2|$
is determined by the lowest-order vector vertex correction, that is, the
middle diagram on the right-hand side of
Fig.~\ref{fig:sv_diag}.\footnote{%
There is also a lowest-order vertex correction involving the exchange
of a scalar meson.
This additional diagram has a range similar to the diagram that we evaluate,
so its inclusion would not significantly change the values that we obtain
for the rms baryon radii.%
}
Then, by knowing the large $|q^2|$ behavior of the form factors, we can
interpolate smoothly between the small and large momentum-transfer regimes.
Note that we need only spacelike (or Euclidean) momenta to compute the energy,
so the form factors should be smooth, which makes the interpolation practical.

The sum of the diagrams for either the scalar or the vector vertex in
Fig.~\ref{fig:sv_diag} results in a rapid suppression at large spacelike
momentum transfers.
As shown by Fishbane and Sullivan\cite{FISHBANE71},
the leading logarithmic
asymptotic behavior for the on-shell vector charge form factor
at zero density can be written as
\begin{equation}
F_{{\rm v}1} (-q^2) \longrightarrow \exp
    \left[F_{{\rm v}1}^{(1)}(-q^2)\right]
    \  , \label{f1asym}
\end{equation}
where $q = p_b - p_a$ and the superscript ``(1)'' indicates the lowest-order
{\em correction}.
Since the anomalous form factor is suppressed asymptotically by an additional
factor of $1/|q^2|$, we can write
\begin{eqnarray}
2MF_{{\rm v}2} (-q^2) &\longrightarrow&
            2MF_{{\rm v}2}^{(1)} (0)\, \nonumber \\
     & & {}\times{1\over |q^2|}\,
         \exp \left[F_{{\rm v}1}^{(1)}(-q^2)\right]
         \ , \label{f2asym}
\end{eqnarray}
where the scale is set by the $O(g _{\rm v}^2 )$ anomalous moment, which is a
conservative choice.

At each order in $g _{\rm v}^2$, the dressed scalar vertex
in Fig.~\ref{fig:sv_diag}
has the same denominator
and the same leading power of $q^2$ in the numerator of the integrand as
the dressed vector vertex.
Since these features determine the asymptotic
behavior\cite{FISHBANE71},
we conclude that the scalar and vector vertices behave similarly,
and we can therefore write
\begin{equation}
F_{\rm s} (-q^2) \longrightarrow \exp \left[F_{\rm s}^{(1)}(-q^2)\right]
    \  , \label{fsasym}
\end{equation}
which again holds for the leading logarithmic behavior.

The lowest-order corrections to the  on-shell vector charge and anomalous
form factors are evaluated in Ref.~\cite{ALLENDES92}, and
we simply quote the results here:
\widetext
\begin{eqnarray}
F_{{\rm v}1}^{(1)}(Q^2) &=&
  -\frac{g _{\rm v}^2}{16\pi^2} \int_0^1 {\rm d}u\ \left\{
    \frac{2\left[2(1-u)-u^2+Q^2(1-u/2)^2\right]}{QS(u)}
         \ln\left[ \frac{S(u)+uQ/2}{S(u)-uQ/2} \right]\right.
          \nonumber \\[5pt]
& &\quad\qquad\qquad\qquad\quad \left.
    {}-2u-\frac{2u\left[2(1-u)-u^2\right]}{u^2+m _{\rm v}^2 (1-u)}
    + \frac{2S(u)}{Q}
         \ln\left[ \frac{S(u)+uQ/2}{S(u)-uQ/2} \right]\right\}
                       \ ,  \label{eq:fv1}\\[6pt]
 2MF_{{\rm v}2}^{(1)}(Q^2)  &= &\frac{g _{\rm v}^2}{4\pi^2}
    \int_0^1 {\rm d}u\
    \frac{u(1-u)}{QS(u)}
    \ln\left[ \frac{S(u) + uQ/2}{S(u) - uQ/2}\right] \ ,\label{eq:fv2}
\end{eqnarray}
where
\begin{equation}
    S(u) \equiv \left[ u^2 + \frac{u^2Q^2}{4} +
       m _{\rm v}^2 (1-u) \right]^{1/2} \ ,
\end{equation}
$Q^2\equiv -q^2/M^2$, $Q \equiv \sqrt{Q^2}$, and $m_{\rm v}$
is written in units of the
baryon mass $M$.

The lowest-order correction to the  on-shell scalar vertex
can be calculated analogously.
Thus
\begin{equation}
 F_{\rm s}^{(1)}(-q^2) =
       ig _{\rm v}^2 \int\!{{\rm d}^4 k \over(2\pi)^4}\,
\gamma^{\mu}G_{\scriptscriptstyle \rm F}^0 (p_b-k)
                   G_{\scriptscriptstyle \rm F}^0 (p_a-k)
                 \gamma^{\nu} D^0_{\mu \nu}(k)+{\rm CTC}   \ ,
\end{equation}
where the right-hand side is understood to be evaluated on shell
(${\rlap/ p}_a ={\rlap/ p}_b = M$), and the counterterm contribution (CTC) is
determined by imposing the on-shell renormalization condition
 $ F_{\rm s}^{(1)} (0) =0 $, which insures that the scalar--baryon coupling
remains $g_{\rm s}$ when the momentum transfer is zero.
Straightforward manipulations
similar to those for the vector vertex\cite{ALLENDES92} yield
\begin{eqnarray}
F_{\rm s} ^{(1)}(Q^2) &=&
       -\frac{g _{\rm v}^2}{16\pi^2} \int_{0}^{1}\!{\rm d}u \left\{
             {12S(u) \over Q} \ln \left[ \frac{S(u) +uQ/2}
                                        {S(u) -uQ/2}\right]
              -12u - \frac{4u(1-u+u^2)}{u^2 +m _{\rm v}^2 (1-u)}\right.
                   \nonumber  \\[5pt]
         & &\qquad\qquad\qquad\quad
         {}+\left. \frac{2 \left[2(1-u+u^2)+Q^2(1-u+\frac{1}{2}u^2)\right]}
              {QS(u)} \ln \left[ \frac{S(u) +uQ/2}
             {S(u) - uQ/2}\right] \right \} \ . \label{eq:fs}
\end{eqnarray}
\narrowtext
It is not hard to show \cite{ALLENDES92}
from Eqs.~(\ref{eq:fv1}) and (\ref{eq:fs}) that for large $Q^2$,
\begin{equation}
  F_{{\rm v}1}^{(1)}(Q^2), \ F_{\rm s} ^{(1)}(Q^2)
            \longrightarrow  - \frac{g _{\rm v}^2}{16\pi^2}
      \ln^2 (Q^2/m _{\rm v}^2)
    \ ,
\end{equation}
to leading order in logarithms.
When combined with Eqs.~(\ref{f1asym}) and (\ref{fsasym}), this agrees with
Eq.~(\ref{eq:basic}).

As mentioned previously, we want to join the low- and
high-momentum transfer regimes using a smooth interpolation.
Evidently, the smaller the region to be interpolated,
the more constrained the interpolation.
On the other hand, the larger
the high-momentum matching point $Q_0$,
the better the leading-logarithm asymptotic behavior.
It is impractical, however, to take $Q_0$ so large that the non-leading
logarithms are negligible compared to the leading logarithms.
Thus, for a trade-off, we will approximate the non-leading
logarithms by the large $Q^2$ behavior of  $F_{{\rm v}1}^{(1)}(Q^2)$
and $F_{\rm s} ^{(1)}(Q^2)$.
That is, we choose $Q_0$ large enough so that
the values of  $F_{{\rm v}1}^{(1)}(Q^2)$
and $F_{\rm s} ^{(1)}(Q^2)$ can be accurately fitted by the function
\begin{eqnarray}
W(Q^2) &=& -\frac{g _{\rm v}^2}{16\pi^2} \left[
\ln ^2(Q^2/m _{\rm v}^2)\right. \nonumber \\
   & &  \qquad\qquad  \left.{}+ r_1\ln (Q^2/m _{\rm v}^2)
   + r_2 \right] \ , \label{r1r2}
\end{eqnarray}
which means that all polynomials in $1/Q^2$ are negligible.
For $Q \geq Q_0$,
we take the vertex form factors to be given by Eqs.~(\ref{f1asym}),
(\ref{f2asym}) and (\ref{fsasym}), with the lowest-order corrections
replaced by Eq.~(\ref{r1r2}).

To minimize the uncertainty in the interpolations, one wants to choose
$Q_0$ as small as is feasible, and we chose $Q_0 = 5$, as in
Ref.~\cite{ALLENDES92}.
The $r_i$ parameters in Eq.~(\ref{r1r2}) were determined by making a
least-squares fit to the expressions in (\ref{eq:fv1}) and
(\ref{eq:fs}).\footnote{%
We note that the onset of the asymptotic regime ($Q_0$) and the values of the
$r_i$ parameters are determined by the behavior of the {\em integrals\/}
in Eqs.~(\protect\ref{eq:fv1}) and (\protect\ref{eq:fs}).
Thus these variables are all independent of the strength of the coupling
$g_{\rm v}$.%
}
To test the sensitivity to the non-leading logarithmic behavior, we obtained
one set of parameters for $5 \leq Q \leq 40$, and then obtained another
(``exact'') set by letting $Q^2 \rightarrow \infty$.
(The exact $r_1$ and $r_2$ can be determined
analytically in principle, but the numerical evaluation is sufficiently
accurate for our purposes.)
Our first least-squares fit yields $r_1
\approx -0.235 $ and $r_2 \approx -1.60$
for the scalar vertex,  and $r_1 \approx -2.88 $ and $ r_2 \approx 5.76$ for
the vector vertex, while the exact results are $r_1 = 0$ and $r_2 = -3.35$ for
the scalar vertex, and $r_1 = -3.00$ and $r_2 = 6.66$ for the vector vertex.
(These values assume $m_{\rm v} = 783\rm\, MeV$ and $M = 939\rm\,
MeV$.)
The resulting interpolations
for the two choices of $r_i$ (as well as for $r_1 = r_2 =
0$) are nearly indistinguishable, which implies a very small sensitivity to
the non-leading logarithmic behavior;
for aesthetic reasons, we will use
the ``exact'' parameters in all interpolations henceforth.

In Ref.~\cite{ALLENDES92},
the last two terms in Eq.~(\ref{r1r2}) were approximated by
introducing a single parameter $\alpha$ such that the asymptotic behavior is
$\exp [ -(g _{\rm v}^2/16\pi^2)
\ln ^2 (Q^2/\alpha m _{\rm v}^2)]$.
For the scalar vertex, this is not as conservative as our present method.
(That is, the method of Ref.~\cite{ALLENDES92} leads to a scalar vertex
function that decays {\em more\/} rapidly with increasing $Q$.)
We therefore use Eqs.~(\ref{f1asym}),
(\ref{f2asym}), and (\ref{fsasym}) for $Q \geq Q_0 = 5$,
with the lowest-order corrections
replaced by Eq.~(\ref{r1r2}).
We take the values of the
form factors and their derivatives at $Q=0$, as given  by their lowest-order
corrections, together with the values and the derivatives at $Q_0$,
determined from the asymptotic forms, as four
input parameters to specify the interpolation functions.

In Ref.~\cite{ALLENDES92}, the interpolating function was taken as
\begin{equation}
f_{\rm M}(Q^2) = a \exp (-bQ^2) + c/(1+Q^2) + d \ , \label{form1}
\end{equation}
where $a$, $b$, $c$, and $d$ are to be determined.
This interpolating function contains both gaussian
and monopole terms, so we will call it a ``mixed''
interpolation for convenience of description.
As noticed in
Ref.~\cite{ALLENDES92},
for some values of $g _{\rm v}^2$, two different solutions for the
parameters $a$, $b$, $c$, and $d$  can be found, while for other values,
unique solutions exist.
Whereas this behavior can
furnish an estimate of the uncertainty in the interpolation, it also
results in a nuclear matter energy that is a discontinuous function
of $g _{\rm v}^2$.
This is inconvenient when one attempts to refit the couplings to
reproduce nuclear matter saturation.

Thus we use instead the following polynomial interpolating function,
\begin{eqnarray}
f_{\rm P}(Q^2) &=& a/(1+Q^2)^3  +
       b/(1+Q^2)^2 \nonumber \\
       & & {}+ c/(1+Q^2) + d \ , \label{form2}
\end{eqnarray}
which yields a unique solution for $a$, $b$, $c$,
and $d$ at any $g _{\rm v}^2$.
To estimate the uncertainties in the interpolation and their effects on the
energy density of nuclear matter, we also use the gaussian
interpolating function
\begin{eqnarray}
f_{\rm G}(Q^2) &=& a \exp (-Q^2)  + b \exp (-Q^2/5)\nonumber \\
     & &{} + c\exp (-Q^2/10) + d
           \ , \label{form3}
\end{eqnarray}
which also yields a unique solution for $a$, $b$, $c$,
and $d$ at any $g _{\rm v}^2$.
Since $F_{{\rm v}2} (q^2)$ is suppressed asymptotically by a factor of
$1/|q^2|$ relative to $F_{{\rm v}1} (q^2)$,
we take  its corresponding interpolating functions
to be those of Eqs.~(\ref{form2}) and (\ref{form3}) multiplied
by $2M F^{(1)}_{{\rm v}2} (0)/(1+Q^2)$.

In Fig.~\ref{fig:sv} we show the
scalar  form factor and the vector charge  form factor using the
mixed, polynomial, and gaussian interpolating functions for RHA couplings.
The vector anomalous form factors obtained from
Eqs.~(\ref{form2}) and (\ref{form3}) with
a further multiplicative
factor of $2M F^{(1)}_{{\rm v}2} (0)/(1+Q^2)$ are shown in
Fig.~\ref{fig:anm_env}(a).
The uncertainty envelopes generated by the  polynomial
and gaussian interpolations are combined in
Fig.~\ref{fig:anm_env}(b).
In Fig.~\ref{fig:mono_di}(a)
we show the vector charge form factor and the
corresponding monopole form factor
[$F_{\rm m}(Q^2)=\Lambda^2/(Q^2+\Lambda^2)$]
that decays similarly, while in
Fig.~\ref{fig:mono_di}(b), we show similar results for a dipole form factor
[$F_{\rm d}(Q^2)=\Lambda^4/(Q^2+\Lambda^2)^2$].
The parameters $a$, $b$, $c$, and $d$ for
the preceding interpolations are listed in
Table~\ref{tab:abcd}.
All values assume $g _{\rm v}^2 = 102.8$, $m_{\rm v} = 783{\,\rm MeV}$,
and $M = 939{\,\rm MeV}$.

The preceding figures indicate that the decay of the form factors
as a function of $Q$ depends significantly
on the interpolation functions.
Nevertheless, we have verified that the polynomial and gaussian forms provide
a reasonable envelope on the uncertainty introduced by the interpolation.
We will therefore use the polynomial and gaussian functions to
investigate the sensitivity of the nuclear matter binding energy to the
decay of the form factors.


\section{RESULTS} \label{sec:four}

We now present numerical results for the vertex-corrected two-loop energy
computed with the equations in Sec.~\ref{sec:two} and vertices in
Sec.~\ref{sec:three}.
We start with the results of a perturbative calculation,
in which the RHA couplings {\em and\/} self-consistent
mass $M^{\scriptstyle\ast}$
are used to compute the two-loop corrections.
For convenience of comparison with the results of
Ref.~\cite{FURNSTAHL89}, we use the couplings $g_{\rm s}^2 = 54.3$ and
$g _{\rm v}^2
= 102.8$, and masses $m_{\rm s} = 458{\,\rm MeV}$
and $m_{\rm v} = 783{\,\rm MeV}$,
which produce nuclear matter
equilibrium at $k_{\scriptscriptstyle\rm F} = 1.30\,{\rm fm}^{-1}$
with a binding energy/nucleon of $15.75{\,\rm MeV}$.
Later we present results obtained by refitting the parameters so that the
total energy, including the two-loop contributions and the new
self-consistent $M^{\scriptstyle\ast}$, reproduces
the preceding equilibrium properties of nuclear matter.
Note that all vertex functions used in this section are fixed at their
free-space parametrizations, which follow from the formulas in
Sec.~\ref{sec:three}.
Calculations both  with and
without the anomalous form factor $F_{{\rm v}2}$ will be considered.

The results for the perturbative
two-loop energy/baryon of nuclear matter as a function
of  density,  
with both the polynomial
and gaussian interpolations for the vertices, are shown in
Fig.~\ref{thesis3.14}.
As a check of our
renormalization procedure and numerical calculations, we reproduced the
two-loop results with bare vertices as
shown in Ref.~\cite{FURNSTAHL89} by setting $F_{\rm s}=F_{{\rm v}1}=1$ and
$F_{{\rm v}2}=0$.\footnote{We can also reproduce the results
in Ref.~\protect\cite{PRAKASH91} by
setting $F_{{\rm v}2}=0$ and choosing $F_{\rm s}$ and $F_{{\rm v}1}$ equal
to the square
of the monopole form factors therein.}
By comparing the results of Fig.~\ref{thesis3.14} with those in
Fig.~4 of Ref.~\cite{FURNSTAHL89},
we observe that the form factors  suppress the vacuum contributions
considerably.
Moreover, the difference between the two interpolations gives an
estimate of the uncertainty in the model vertex functions, which produces
large changes on the scale of the nuclear binding energy.
Thus the energy density of nuclear matter in this model
is sensitive to the behavior of
the form factors at intermediate momenta,
at least in the two-loop approximation.

The separate contributions
from the scalar, the vector charge ($F_{{\rm v}1}$),
and the vector anomalous ($F_{{\rm v}2}$) parts are
shown in Fig.~\ref{thesis3.15}.
Evidently, there are strong cancellations between the vector charge and
scalar pieces.
Notice also that the anomalous
contributions are generally only $5\%$--$15\%$ as large as the scalar
and vector charge contributions.
Nevertheless, the anomalous terms are
comparable to the  binding energy of nuclear matter and are
relevant in a calculation of nuclear matter saturation because of
the sensitive cancellations between the vector charge and scalar
contributions.

The separate contributions to the two-loop correction from
the vacuum fluctuation, the exchange, and the Lamb-shift terms are
shown
in Fig.~\ref{thesis3.16} with $F_{{\rm v}2}$ omitted
and in Fig.~\ref{thesis3.17} with $F_{{\rm v}2}$ included.
One observes that the Lamb-shift energy suffers less suppression from
the form factors than the vacuum fluctuation energy does.
When $F_{{\rm v}2}$ is included, the vacuum fluctuation and Lamb-shift
energies
become more negative.

The two-loop energy with vertex modifications
${\cal E}^{(1)}+{\cal E}^{(2)}$
can be minimized with respect to
$M^{\scriptstyle\ast}$ at each density, and the parameters can be
adjusted to reproduce
the equilibrium properties of nuclear matter.
The vector meson mass is held fixed at its empirical value of
$783{\,\rm MeV}$ (for simplicity),
but the scalar meson mass must be increased, with the
size of the increase determined by
how big the two-loop perturbative result deviates from the RHA result.
(If too small a value is chosen for the scalar mass, the resulting
interaction is too attractive, and nuclear matter always saturates with
too large a binding energy.)

We have found a number of parameter sets that reproduce the equilibrium
properties of nuclear matter.
A representative sample is listed in Table~\ref{tab:paras}, where
the compressibility $K^{-1}_{\rm v}$ is also shown.\footnote{%
As is well known, the compressibility can be reduced, if desired, by
including nonlinear scalar meson self-interactions.}
Fig.~\ref{thesis3.18} shows the corresponding refitted
results with and without the anomalous
form factor and with the polynomial or the gaussian interpolation.
In general, the compressibility remains large in this approximation and
increases slightly from the value in the RHA; the form factors with the
polynomial interpolation produce more suppression of the vacuum contributions
(compared to the gaussian interpolation), but give a higher compressibility.
The corresponding self-consistent nucleon mass
is shown as a function of density in Fig.~\ref{thesis3.19}.
Here $M^{\scriptstyle\ast}$ is obtained by minimizing the full
two-loop energy density
${\cal E}^{(1)}(M^{\scriptstyle\ast}) +
{\cal E}^{(2)}(M^{\scriptstyle\ast})$ with respect to this parameter.
The two-loop contributions generally reduce the value of
$M^{\scriptstyle\ast}$.

The RHA and two-loop contributions to the total
refitted energy with $F_{{\rm v}2}$ excluded are shown in
Fig.~\ref{thesis3.20},
where we also show the
vacuum fluctuation, the exchange, and the Lamb-shift energies.
The RHA  and two-loop contributions to the total
refitted energy with $F_{{\rm v}2}$ included are shown in
Fig.~\ref{thesis3.21},
where the scalar, vector charge, and vector anomalous parts in the
two-loop correction are also shown.

To refit the energy/baryon to nuclear-matter equilibrium,
the scalar mass must be increased significantly, which
suggests that the results in this approximation are
sensitive to the precise shape of the form factors.
Nevertheless, the two-loop calculation with vertex
modifications is a better approximation than the conventional
two-loop calculation with bare vertices, since the two-loop corrections
are smaller.
For quantitative comparisons to nuclear matter properties, one must
know the form factors more accurately, especially in the intermediate
momentum transfer region.
Thus our calculations imply that
it is inappropriate to draw conclusions from calculations using some
particular {\em ad hoc} choice of
form factors, as in Refs.~\cite{PRAKASH91} and
\cite{FRIEDRICH92}.

%

\section{RESULTS WITH MEDIUM-MODIFIED VERTICES} \label{sec:five}

One advantage of having an explicit model for the meson--baryon vertex
functions is that we can investigate how these functions change in the nuclear
medium.
For example, the baryon mass changes from $M$ to
$M^{\scriptstyle\ast}$, and by making this
change in the baryon propagators that appear in the vertex loops, the vertex
functions acquire an implicit density dependence.
This is the effect we will consider in this section.
We emphasize that this is a first step in the study of density-dependent
vertex modifications in this model,
since we are neglecting the valence-nucleon
contributions to the vertex loops,
as well as additional form factor functions that can arise
at finite density.
Moreover, as with the calculations in the preceding sections, a quantitative
study requires the full off-shell vertex, as well as the inclusion of
pions to more accurately describe the long-range vertex structure.
We leave these additional modifications as important topics for future
investigations.
Although the present calculation is just a first step, it illustrates
numerous issues that must be dealt with in any microscopic model of the
meson--baryon vertices; many of these issues are simply omitted by assumption
in conventional calculations based on {\em ad hoc\/} form factors.

By replacing $M$ with $M^{\scriptstyle\ast}$ in the vertex
functions, the mass of the
virtual intermediate state is reduced, and thus the radius of the
dressed nucleon increases with increasing density.
This is the primary modification we study here.
We emphasize that the leading logarithmic behavior at large spacelike $Q^2$ is
unchanged.
However, since we determine the non-leading behavior from
an examination of the
lowest-order loop diagrams, this behavior also changes at finite density;
these effects are very small and are incorporated here just for
consistency.
We will follow the general strategy described earlier: include effects that
can be calculated reliably in our simple model, and treat effects that
cannot yet be calculated as conservatively as possible.

It is a straightforward exercise to generalize the results in
Sec.~\ref{sec:two} to
incorporate the modified baryon mass, and we simply quote the results here:
\widetext
\begin{eqnarray}
F_{\rm s} ^{\ast(1)}(Q^2) &=& -\frac{g _{\rm v}^2}{16\pi^2}
             \int_{0}^{1}\!{\rm d}u \left\{
             12S^{\ast}(u) L(u)
              -12u - \frac{4u(1-u+u^2)}{u^2 +m _{\rm v}^2 (1-u)}\right.
                   \nonumber  \\[5pt]
         & &\quad\qquad\qquad\qquad {}+ \frac{2
             \left[2m^{\scriptstyle \ast}{}^2(1-u+u^2)
                +Q^2(1-u+\frac{1}{2}u^2)\right]}
              {S^{\ast}(u)} L(u)
                    \nonumber \\[5pt]
         & & \quad\qquad\qquad\qquad {}+\left. 8u \ln
            \left[ \frac{m^{\scriptstyle \ast}{}^2 u^2 + m _{\rm v}^2 (1-u)}
                                  {u^2 + m _{\rm v}^2 (1-u)} \right]
             \right\} \ , \label{eq:fsstar}\\[7pt]
F_{{\rm v}1}^{\ast(1)}(Q^2) &=&
    -\frac{g _{\rm v}^2}{16\pi^2} \int_0^1 {\rm d}u \left\{
    \frac{2\left\{m^{\scriptstyle \ast}{}^2\left[2(1-u)-u^2\right]
        +Q^2(1-u/2)^2\right\} }
            {S^{\ast}(u)} L(u)\right.
          \nonumber \\[5pt]
& &\quad\qquad\qquad\qquad
    {}-2u-\frac{2u\left[2(1-u)-u^2\right]}{u^2+m _{\rm v}^2 (1-u)}
    + 2S^{\ast}(u) L(u)
        \nonumber \\[5pt]
& &\quad\qquad\qquad\qquad \left.
         {} + 2u \ln \left[ \frac{m^{\scriptstyle \ast}{}^2 u^2
                + m _{\rm v}^2 (1-u)}
                      {u^2 + m _{\rm v}^2 (1-u)} \right]
                    \right\}
                       \ ,  \label{eq:fv1star}\\[7pt]
 2MF_{{\rm v}2}^{\ast(1)}(Q^2)  &= &\frac{g _{\rm v}^2}{4\pi^2}
    \int_0^1 {\rm d}u\
    \frac{m^{\scriptstyle \ast} u(1-u)}{S^{\ast}(u)} L(u)
        \ ,\label{eq:fv2star}
\end{eqnarray}
\narrowtext
\noindent
where
\begin{eqnarray}
    L(u) &\equiv& {1 \over Q}\,
    \ln\left[ \frac{S^{\ast}(u) + uQ/2}{S^{\ast}(u) - uQ/2}\right]
        \ , \\[5pt]
    S^{\ast}(u) &\equiv& \left[ m^{\scriptstyle \ast}{}^2 u^2 +
          \frac{u^2Q^2}{4} + m _{\rm v}^2 (1-u) \right]^{1/2} \ ,
\end{eqnarray}
$Q^2\equiv -q^2/M^2$, $Q \equiv \sqrt{Q^2}$,
$m^{\scriptstyle \ast} \equiv M^{\scriptstyle\ast} /M$,
and $m_{\rm v}$ is written in units of the baryon mass $M$.
We note the important point that the prefactor $2M$ is used solely to make
the anomalous contribution dimensionless (any mass could be used), and thus
it does {\em not\/} change to $M^{\scriptstyle\ast}$ at finite density.

Here we have renormalized the form factors using the same
renormalization conditions as in Sec.~\ref{sec:three}.
A straightforward numerical evaluation of the preceding integrals
shows that they do not vanish at $Q^2 = 0$, and thus the vector charge and
scalar couplings are renormalized at finite density.
This is especially surprising for the vector term, since the conservation of
the baryon current implies that there is no charge renormalization.
The resolution of this dilemma is simple.
In a calculation that includes the full off-shell vertex inside
the two-loop diagrams, one
should also include the appropriate self-energy insertions on the baryon
lines to maintain the required Ward identities.
Thus, when we ``extract'' the vertex and renormalize it on mass-shell, using
the procedures discussed in Sec.~\ref{sec:two},
we should also extract the self-energies,
which now appear on external baryon lines.
In free space, these self-energy
insertions vanish by construction, due to our choice of
mass and wave-function counterterms, but when $M$ is
replaced by $M^{\scriptstyle\ast}$,
one finds a finite wave-function renormalization.
It is straightforward to show\cite{SCHWEBER61}
that the contribution from this wave-function
renormalization precisely cancels the shift from
$F_{{\rm v}1}^{\ast(1)}(0)$, and the baryon number remains unchanged.
Thus we will set the vector charge form factor
$F^{\ast}_{{\rm v}1} (0) = 1$ in
performing the interpolations discussed below.

The self-energy insertions also produce a finite
correction to the baryon mass, which is exactly what
is needed to modify the mass in the external spinor to include the correction
from the exchange self-energy.
Since we will only consider perturbative calculations about the RHA results,
in which we neglect all exchange corrections to the baryon mass (by
definition), we will neglect this mass shift as well.

In contrast, the scalar vertex is not protected by the Ward identity.
The finite wave-function renormalization does not cancel exactly against
$F_{\rm s} ^{\ast(1)}(0)$,
and there is a finite shift in the strength of the vertex.
However, while we believe that the lowest-order vertex correction gives a
reasonable estimate of the form-factor radius (since it incorporates the
intermediate state with the lowest mass),
there is no reason to expect that it
is reliable for the renormalization of the scalar coupling strength.
This renormalization is a completely dynamical effect that is likely to
depend significantly on valence-nucleon contributions, scalar meson exchange
diagrams, vacuum polarization, {\em etc.}, all of which are neglected here.
We will therefore take a conservative approach and leave the magnitude of the
scalar strength {\em unrenormalized\/} at finite density; we postpone
the difficult problem of obtaining a reasonable estimate for the finite
renormalization to a future investigation.

It is also straightforward to obtain expressions for the medium-modified rms
radii of the nucleon, but we will not present the formulas here.
(These are most easily obtained by differentiating the Feynman parameter
integrals with respect to $Q^2$,
rather than by differentiating the expressions
given above.)
One finds that the vector charge and scalar mean-square radii
scale approximately as
$1/M^{\scriptstyle\ast}$.
Because of the overall factor
of $m^{\scriptstyle \ast}$ in $2MF^{\ast}_{{\rm v}2}$, however, the
anomalous moment (and its radius) are
insensitive to the value of $M^{\scriptstyle\ast}$,
and this form factor is essentially the same as at zero density.

The parametrizations of the vertex functions are performed as described in
Sec.~\ref{sec:three}, with a different parameter set
at each value of $M^{\scriptstyle\ast}$.
One finds that the values of $r_1$ [see Eq.~(\ref{r1r2})]
are unchanged at finite density, while the values of $r_2$ suffer a (small)
change that is exactly the same for the scalar and vector vertex.
Fig.~\ref{vecmstar} shows the vector charge form factor at
$M^{\scriptstyle\ast}=0.7M$, where we have included the finite wave-function
renormalization, so that $F^{\ast}_{{\rm v}1} (0)=1$.

If one faithfully carries out the renormalization procedure defined in
Sec.~\ref{sec:two} using the medium-modified vertices,
one discovers that the resulting
expressions for the two-loop energy are no longer finite.
This occurs because the two-loop integrals contain the modified vertices,
while the subtraction terms are defined by vacuum amplitudes and use
unmodified vertices; thus, certain divergences no longer cancel.
The problems can be traced to the nested divergences, which require the
renormalization of the self-energy and vertex functions, as discussed in
the Appendix.
(The overlapping subtractions remain unchanged, and the overall subtractions
are always chosen to remove
the first four powers of $\phi_{\scriptscriptstyle 0}$.)
In principle, the only way to determine the required new subtractions is to
use the full off-shell vertices inside the loops.
Within the context of the on-shell approximation for the form factors,
{\em the best that can be done\/} is to use medium-modified vertices for both
the two-loop integrals {\em and\/} the corresponding subtractions.
This is the procedure we follow here.
The result is that the energy density is calculated by inserting the
density-dependent vertex functions, parametrized as discussed earlier,
into the expressions for the renormalized
two-loop contributions given in Sec.~\ref{sec:two}.

In Fig.~\ref{en_2lp_mst}, we show the
results of a perturbative calculation (RHA
parameters and $M^{\scriptstyle\ast}$)
using the medium-dependent vertices with both the
polynomial and gaussian interpolations.
A comparison with Fig.~\ref{thesis3.14} shows that the two-loop
corrections are
smaller here, as expected from the larger radii (and consequent
more rapid decay in momentum space) of the $F^{\ast}_{{\rm v}1}$ and
$F^{\ast}_{\rm s}$ vertices.
The differences between the results with the two different interpolations is
again an estimate of the uncertainty in the form factors at
intermediate momenta; these differences remain
significant on the scale of the nuclear matter binding energy.
Moreover, it is clear that including
the density dependence of the vertices also produces
effects that are significant on this scale.
This suggests that for quantitative calculations of nuclear matter
saturation, it is necessary to include this density dependence, and
one can question the accuracy of results using vertices that are simply
assumed to be density independent.

\section{SUMMARY}

We have calculated the two-loop energy with one dressed vertex for
nuclear matter in the Walecka model, using form factors determined in
the framework of the model.
With an on-shell approximation for the vertex,
we determine the leading behavior at large spacelike momentum transfer from
the sum of an infinite set of ladder and crossed-ladder diagrams, and
we use the lowest-order vertex correction to specify the behavior at small
momentum transfer.
Two different interpolation functions are used to join the two
momentum-transfer regimes.
Whereas the resulting vertex functions are not determined uniquely at
intermediate momentum transfers and are known only to within an error band,
they are specified by the underlying couplings and masses in the model
without the introduction of {\em ad hoc\/} parameters.

We find that the two-loop corrections to the one-loop energy are considerably
smaller than those computed earlier with bare vertices.
While these corrections produce significant changes in the nuclear binding
energy, one can still fit the empirical equilibrium properties of nuclear
matter with a reasonable adjustment of the model parameters.
(The most dramatic adjustment is in the scalar meson mass, which must be
increased by roughly several hundred MeV.)
Since these results were obtained using only one dressed vertex in the loop,
it is likely that calculations using two dressed vertices will find even
smaller corrections.

We also included the contributions from the anomalous vector vertex for
the first time.
Although its contribution is
small compared to those of the vector charge vertex or the scalar vertex,
it is not negligible on the scale of the nuclear matter binding energy.
Moreover, the implicit density dependence in the vertices and
the uncertainty in our interpolations also produce variations that are
important on the scale of the binding energy.
These new effects should be present in any realistic model of the
meson--baryon vertex functions and must be included before a quantitative
calculation of nuclear matter saturation can be performed.

Although the present
calculations are the first step in the inclusion of dynamically
generated vertex functions in relativistic nuclear matter calculations, and
many improvements and refinements must be made, there are several features
of our analysis that we believe to be under control.
First, the leading logarithmic behavior of the vertices at large spacelike
momentum transfer
is included correctly.
Second, the low-momentum-transfer behavior should
be reproduced reasonably well
(for a model with only heavy mesons), since we include the intermediate state
with the smallest mass.
Third, our two different interpolation functions give a realistic ``error
envelope'' for the uncertainty of the vertex functions at intermediate
momentum transfer and for the resulting uncertainty in the nuclear
matter binding energy.
Fourth, the calculated isoscalar anomalous moment has a magnitude that is
consistent with empirical values, and thus it provides a meaningful
estimate of the size of the isoscalar anomalous contributions to nuclear
matter saturation.
Finally, the mean-square radii of the
vector charge and scalar vertices increase with
density  roughly as the inverse of the baryon effective mass, which should
give sensible results for the impact of this density dependence on the
nuclear matter calculation.

There are several improvements that must be made in these calculations before
any definitive conclusions can be drawn.
First, one must compute the off-shell vertices and include them inside the
loop integrals; this will also require a computation of the corresponding
self-energies, which must be included to maintain the conservation of
the baryon current\cite{ALLENDES93}.
Second, the contributions from the density-dependent (valence) parts of the
baryon propagators must be included in the computation of the vertex
functions; this will introduce (in principle) new vertex functions that
arise at finite density.
Third, pions must be included to describe the long-range vertex structure
more accurately.
Fourth, the description of the vertex functions at intermediate momentum
transfer should be improved; this may be possible by examining the
dispersion relations that determine these functions.
Finally, one must investigate the truncation procedure used to define the
present approximation to decide if it is more accurate to dress one or both
of the vertices in the two-loop terms.
All of these improvements can be studied systematically within the QHD
framework and provide topics for future work on this problem.

In a larger context, regardless of which degrees of freedom one believes are
the most appropriate, it is necessary to have reliable models of the off-shell
and density-dependent behavior of the meson--baryon vertices before
accurate relativistic calculations of nuclear matter properties can be made.
The construction of such models presents a formidable challenge to the
practitioners of relativistic nuclear many-body theory.

\acknowledgments

We are pleased to thank P. J. Ellis, R. J. Furnstahl, C. J. Horowitz,
J. Milana, and R. J. Perry for their useful comments.
We are also grateful to S. V. Gardner for a careful reading of a draft of
this paper.
This work was supported in part by the U.S. Department of Energy under
contract No.~DE--FG02--87ER40365 and by the National Science Foundation
under Grant Nos.~PHY-9203145, PHY-9258270, PHY-9207889, and PHY-9102922.

\newpage
\appendix

\section*{BARYON SELF-ENERGY AND 
BARYON--SCALAR VERTEX}

Here we derive the renormalized
Feynman parts of the baryon self-energy and baryon--scalar vertex
that appear in Eqs.~(\ref{eq:slf}) and (\ref{eq:vtx}).
The baryon self-energy is shown graphically in
Fig.~\ref{fig:sigma}.
The baryon--scalar vertex is shown in
Fig.~\ref{fig:bsv}, with the counterterm subtraction omitted.
Here we will be interested only in the case
of zero momentum transfer.
Our  renormalization procedure is similar to that in Ref.~\cite{LIM90}.

It follows from Eq.~(\ref{eq:slf}) that
\begin{equation}
\Sigma_{\scriptscriptstyle\rm F} (k)
= \Sigma_{\scriptscriptstyle\rm FA} (-k^2)
+\rlap/{\mkern-1mu k} \Sigma_{\scriptscriptstyle\rm FB} (-k^2)
                        \ , \label{appdx:slf}
\end{equation}
with
\widetext
\begin{eqnarray}
\Sigma_{\scriptscriptstyle\rm FA} (-k^2) &= &
            i \int\!{{\rm d}^4 q \over (2\pi)^4}\, \frac{1}
                {(k-q)^2-{M^{\scriptstyle\ast}}^2 + i \epsilon}
           \Big\{g_{\rm s}^2\Delta^0(q)
               M^{\scriptstyle\ast} F_{\rm s}(-q^2)-g _{\rm v}^2
         D^0(q)\nonumber \\
             & &{}\times \left[4M^{\scriptstyle\ast} F_{{\rm v}1} (-q^2)
                 +3(q^2-k\cdot q)F_{{\rm v}2} (-q^2)\right]\Big\}
                  -M\zeta_{\scriptscriptstyle\rm N}-\gamma_{\rm s}
              \phi_{\scriptscriptstyle 0}-M_c
                \ ,\label{sfa} \\
\Sigma_{\scriptscriptstyle\rm FB} (-k^2) &= &
            {i \over k^2}\int\!{{\rm d}^4 q \over (2\pi)^4}\,
                \frac{1} {(k-q)^2-{M^{\scriptstyle\ast}}^2 + i \epsilon}
                \Big \{ \Big[g_{\rm s}^2\Delta^0(q)F_{\rm s}(-q^2)
                \nonumber \\
      & & {}+2 g _{\rm v}^2 D^0(q)F_{{\rm v}1} (-q^2) \Big](k^2-k\cdot q)
               - 3g _{\rm v}^2M^{\scriptstyle\ast} D^0(q)
             F_{{\rm v}2} (-q^2) k\cdot q \Big \}
              +\zeta_{\scriptscriptstyle\rm N} \ . \nonumber \\
              & & \label{sfb}
\end{eqnarray}
The convergence properties of these integrals depend on the behavior of the
form factors; for bare vertices, the integrals are logarithmically divergent.
To renormalize them, we write them in terms of Euclidean momenta and then
explicitly subtract the integrands using counterterms chosen to reproduce the
appropriate renormalization conditions.
(This method is discussed in Refs.~\cite{LIM90} and \cite{TANG93}.)
One can verify that, for bare vertices, this procedure reproduces the results
obtained with the more conventional dimensional
regularization\cite{FURNSTAHL89}, and it allows us to extend the
renormalization to include our model vertex functions.

After making a Wick rotation to Euclidean space
($p_0 \rightarrow ip^{\scriptscriptstyle \rm E}_4$)
and
performing the angular integrals, we can write Eqs.~(\ref{sfa})
and (\ref{sfb}) as
\begin{eqnarray}
\Sigma_{\scriptscriptstyle\rm FA}(s) & = &
            -\frac{1}{64\pi^2s}\int_{0}^{\infty} {\rm d}t \,
                \left\{2[M^{\scriptstyle\ast} A_1(t) +tB(t)]
            \Theta_1(s,t;M^{\scriptstyle\ast}{}^2)
                     \right. \nonumber \\
            & &\quad \qquad\qquad\qquad \ \ \left.
           {}-B(t)\Phi_1(s,t;M^{\scriptstyle\ast}{}^2)\right\}-
                 M\zeta_{\scriptscriptstyle\rm N}-\gamma_{\rm s}
              \phi_{\scriptscriptstyle 0}-M_c
                    \ ,\label{sfa:more} \\
\Sigma_{\scriptscriptstyle\rm FB} (s) & = &
            -\frac{1}{64\pi^2s^2}\int_{0}^{\infty} {\rm d}t
             \left\{2 sA_2(t)\Theta_1(s,t;M^{\scriptstyle\ast}{}^2)
              \right. \nonumber \\
            & &\quad \qquad\qquad\qquad \ \ \left.
           {}-\left[A_2(t)+M^{\scriptstyle\ast} B(t)\right]
                \Phi_1(s,t;M^{\scriptstyle\ast}{}^2)
              \right\} +\zeta_{\scriptscriptstyle\rm N} \ , \label{sfb:more}
\end{eqnarray}
where $A_1(t), A_2(t)$, and $B(t)$ are defined in Eqs.~(\ref{eq:a1}),
(\ref{eq:a2}), and (\ref{eq:bs}), and
\begin{eqnarray}
\Theta_1(s,t;m^2) & = & s+t+m^2 -\left[(s+t+m^2)^2-4st\right]^{1/2} \ , \\
\Phi_1(s,t;m^2) & = & (s+t+m^2)^2-2st-(s+t+m^2)
                \left[(s+t+m^2)^2-4st\right] ^{1/2} \ .
\end{eqnarray}
Here
we have set $s={k_{\scriptscriptstyle\rm E}} ^2$
and $t={q_{\scriptscriptstyle\rm E}} ^2$, with the subscript ``E'' denoting
Euclidean.

Similarly, from Eq.~(\ref{eq:vtx}), we have
\begin{equation}
\Lambda^0(k) =
        \Lambda^0_{\scriptscriptstyle\rm A} (-k^2)
      +\rlap/{\mkern-1mu k} \Lambda^0_{\scriptscriptstyle\rm B} (-k^2)
                        \ , \label{appdx:vtx}
\end{equation}
with
\begin{eqnarray}
\Lambda^0_{\scriptscriptstyle\rm A}(s) & = & -\int_{0}^{\infty} {\rm d}t \,
           \frac{ A_1(t)}{32\pi^2s}\left[\Theta_1(s,t;M^2)+2M^2
       \Theta_2(s,t;M^2)
           \right] \nonumber \\[4pt]
          & & \quad -\int_{0}^{\infty} {\rm d}t \,
           \frac{ M B(t)}{16\pi^2s}\left[t\Theta_2(s,t;M^2)-\Phi_2(s,t;M^2)
            \right]+\frac{\gamma_{\rm s}}{g_{\rm s}}\ , \label{vta} \\[6pt]
\Lambda^0_{\scriptscriptstyle\rm B}(s) & = &- \int_{0}^{\infty} {\rm d}t \,
        \frac{M A_2(t)}{16\pi^2s^2}\left[ s\Theta_2(s,t;M^2)-\Phi_2(s,t;M^2)
          \right] \nonumber \\[4pt]
          & & \quad +\int_{0}^{\infty} {\rm d}t \,
           \frac{ B(t)}{64\pi^2s^2}\left[\Phi_1(s,t;M^2)+4M^2\Phi_2(s,t;M^2)
            \right]\ ,
\end{eqnarray}
where
\begin{eqnarray}
\Theta_2(s,t;m^2) & = &1- {s+t+m^2 \over \left[(s+t+m^2)^2-4st\right]^{1/2}}
                                                   \ , \\
\Phi_2(s,t;m^2) & =& s+t+m^2-{(s+t+m^2)^2-2st \over
                \left[(s+t+m^2)^2-4st\right] ^{1/2}} \ .
\end{eqnarray}

The counterterm contributions in Eqs.~(\ref{sfa:more}), (\ref{sfb:more}),
and (\ref{vta})
are defined by imposing the usual conditions
\begin{eqnarray}
\left. \Sigma_{\scriptscriptstyle\rm F}(p) \right|_{\not p = M =
M^{\scriptstyle\ast}}
        & = & 0 \ ,             \label{RR16} \\
\left. {\partial \over\partial {\rlap/ p}}
    \Sigma_{\scriptscriptstyle\rm F}(p)
    \right|_{\not p = M = M^{\scriptstyle\ast}} & = & 0  \ , \\[4pt]
\left. \Lambda ^0(p) \right|_{\not p = M} & = & 0   \label{RR18} \ .
\end{eqnarray}
After inserting the expresions for the self-energy and vertex
[Eqs.~(\ref {appdx:slf}) and (\ref {appdx:vtx})] into
Eqs.~(\ref {RR16})--(\ref {RR18}), it is
straightforward to find the counterterm contributions and to make the
necessary subtractions.
After some tedious algebra, the final results are
\begin{eqnarray}
\Sigma_{\scriptscriptstyle\rm {FA}}(s) & = & - \int_{0}^{\infty}\!\!
          {\rm d}t\,
          \frac{A_1(t)}{32\pi^2}\left\{ {m^{\scriptstyle \ast} \over s}
          \Theta_1(s,t;m^{\scriptstyle \ast}{}^2)
       {}+(2+m^{\scriptstyle \ast})t+2m^{\scriptstyle \ast}-{1\over \eta_t}
           \left[(2+m^{\scriptstyle \ast})t
        +2(2+3m^{\scriptstyle \ast})\right]\right\} \nonumber \\
   & &-  \int_{0}^{\infty}\!\! {\rm d}t\, \frac{A_2(t)}{16\pi^2}
         \left\{t^2+(2+m^{\scriptstyle \ast})t+m^{\scriptstyle \ast}
             -{1\over \eta_t}\left[t^2+(4+m^{\scriptstyle \ast})t
            +2+3m^{\scriptstyle \ast}\right]
             \right\}    \nonumber \\
   & &
   - \int_{0}^{\infty}\!\! {\rm d}t \,\frac{B(t)}{64\pi^2}
          \left\{ {2t\over s}\Theta_1(s,t;m^{\scriptstyle \ast}{}^2)-
                     {1\over s}\Phi_1 (s,t;m^{\scriptstyle \ast}{}^2)
        +t[(6+m^{\scriptstyle \ast})t+6m^{\scriptstyle \ast}]
                 \right. \nonumber \\
   & &\qquad\qquad\qquad\qquad \left.
              {}-{t\over \eta_t}[(6+m^{\scriptstyle \ast})t
         +8(2+m^{\scriptstyle \ast})] \right\}    \ ,\\ [8pt]
\Sigma_{\scriptscriptstyle\rm {FB}}(s) & = & \int_{0}^{\infty}\!\!  {\rm d}t\,
         \frac{A_1(t)}{16\pi^2}\left[t+1-{1\over \eta_t}(t+3)\right]
                                  \nonumber \\
    & &-\int_{0}^{\infty}\!\! {\rm d}t \,
             \frac{A_2(t)}{64\pi^2}
           \left\{ {2\over s}\Theta_1(s,t;m^{\scriptstyle \ast}{}^2)
                -{1\over s^2}\Phi_1(s,t;m^{\scriptstyle \ast}{}^2)
            {}-(3t^2+8t+4)+{1\over \eta_t}(3t^2+14t+12) \right\}
                                 \nonumber \\
    & &+\int_{0}^{\infty}\!\! {\rm d}t \,
             \frac{B(t)}{64\pi^2}
           \left\{{m^{\scriptstyle \ast} \over s^2}
           \Phi_1(s,t;m^{\scriptstyle \ast}{}^2)+
          5t^2+6t-{t\over \eta_t}(5t+16)  \right\}\ , \\ [8pt]
\Lambda^0_{\scriptscriptstyle\rm A}(s) & = &-\int_{0}^{\infty}\!\!  {\rm d}t\,
            \frac{A_1(t)}{32\pi^2}\left\{{1\over s}\Theta_1(s,t;1)+
                 {2\over s}\Theta_2(s,t;1)
            +t+2-{1\over \eta_t}(t+6)\right\} \nonumber \\
   & &  -\int_{0}^{\infty} \!\! {\rm d}t \,\frac{A_2(t)}{16\pi^2}
         \left[t+1-{1\over \eta_t}(t+3)\right] \nonumber \\
   & & -\int_{0}^{\infty}\!\!  {\rm d}t\,
            \frac{B(t)}{16\pi^2s}\left\{t\Theta_2(s,t;1)-\Phi_2(s,t;1)
            +\frac{st}{4}\left[t+6-{1\over \eta_t}(t+8)\right]\right\}
                    \ ,\\ [8pt]
\Lambda^0_{\scriptscriptstyle \rm B}(s) & = &-\int_{0}^{\infty}\!\!{\rm d}t\,
            \frac{A_2(t)}{16\pi^2s^2}[s\Theta_2(s,t;1)-\Phi_2(s,t;1)]
   +\int_{0}^{\infty}\!\!  {\rm d}t\,
            \frac{B(t)}{64\pi^2s^2}[\Phi_1(s,t;1)+4\Phi_2(s,t;1)] \ ,
\end{eqnarray}
\narrowtext
\noindent
where we have scaled all dimensional variables with the nucleon mass
$M$ and defined
$\eta_t \equiv \sqrt{1+4/t}$~
and $m^{\scriptstyle \ast} \equiv M^{\scriptstyle \ast}/M$.

All of the preceding
integrals can be evaluated using Gaussian quadrature, although care must be
used due to cancellations at large $t$.
It is easily verified
that the integrals that contain $F_{\rm s}$ and $F_{{\rm v}1}$
are all finite even
after setting $F_{\rm s}=1$ and $F_{{\rm v}1} =1$,
while the integrals  that contain
$F_{{\rm v}2}$ are also finite when one realizes that in
any realistic calculation,
$F_{{\rm v}2} (s) / F_{{\rm v}1} (s)$ is suppressed
by at least a factor of $1/s$ at large $s$.
We have verified numerically that our results
with $F_{\rm s}=F_{{\rm v}1}=1$ and $F_{{\rm v}2} =0$ are the same
as those of Ref.~\cite{FURNSTAHL89}.
%


 \begin{figure}
 \caption{
 The two-loop corrections to the RHA energy density.
 The solid, dashed,
 and wavy lines represent the baryon Hartree propagator,
 the free scalar propagator, and the free vector propagator,
respectively.
 The subtractions for the overall divergences and for
 the vacuum expectation value of the energy are not shown.}
 \label{FM2loop}
 \end{figure}
 \begin{figure}
 \caption{
 The proper baryon self-energy.
 The diagrammatic notation is the same as in
 Fig.~\protect\ref{FM2loop}.}
 \label{fig:sigma}
 \end{figure}

 \begin{figure}
 \caption{ The unrenormalized two-loop diagrams.}
 \label{2lp_fig}
 \end{figure}

\begin{figure}
\caption{
Diagrammatic expansion of the proper scalar and vector vertices.}
\label{fig:sv_diag}
\end{figure}

\begin{figure}
\caption{
The on-shell scalar form factor (a) and vector charge form factor (b)
with the mixed, polynomial, and gaussian interpolations.}
\label{fig:sv}
\end{figure}

\begin{figure}
\caption{
The on-shell vector anomalous form factor (a) and
the uncertainty envelopes generated by the
gaussian and polynomial interpolations (b). }
\label{fig:anm_env}
\end{figure}

\begin{figure}
\caption{
Comparison of the on-shell vector charge form factor of
Fig.~\protect\ref{fig:sv}(b) with some similar monopole and
dipole form factors. }
\label{fig:mono_di}
\end{figure}

\begin{figure}
\caption{
The perturbative two-loop energy with vertex corrections
using the polynomial and gaussian interpolations. The dotted
(dashed)
curves are obtained with $F_{{\rm v}2}$ excluded (included).
The solid curve is the RHA energy.
Both calculations use RHA parameters and $m_{\rm v} = 783{\,\rm MeV}$.
The equilibrium density $\rho_{\scriptscriptstyle 0}$ corresponds to
$k_{\scriptscriptstyle\rm F} = 1.30\,{\rm fm}^{-1}$.}
\label{thesis3.14}
\end{figure}

\begin{figure}
\caption{
The separate contributions to the total two-loop correction
from the scalar, vector charge, and vector anomalous parts.
The parameters are the same as those in Fig.~\protect\ref{thesis3.14}.
}
\label{thesis3.15}
\end{figure}

\begin{figure}
\caption{
The separate contributions to the total two-loop correction
from the vacuum fluctuation, the exchange, and the Lamb-shift terms,
excluding $F_{{\rm v}2}$. The parameters are the same as those in
Fig.~\protect\ref{thesis3.14}. }
\label{thesis3.16}
\end{figure}

\begin{figure}
\caption{
The separate contributions to the total two-loop correction
from the vacuum fluctuation, the exchange, and the Lamb-shift terms,
including $F_{{\rm v}2}$. The parameters are the same as those in
Fig.~\protect\ref{thesis3.14}. }
\label{thesis3.17}
\end{figure}

\begin{figure}
\caption{
The refitted two-loop energy with vertex modifications
obtained for the polynomial and gaussian
interpolations, respectively. The parameter sets
correspond to those in Table~\protect\ref{tab:paras}.}
\label{thesis3.18}
\end{figure}

\begin{figure}
\caption{
The effective nucleon mass for the
refitted two-loop approximation with vertex modifications
obtained for the polynomial and gaussian interpolations,
respectively.}
\label{thesis3.19}
\end{figure}

\begin{figure}
\caption{
The RHA and the two-loop contributions to the refitted energy density
with $F_{{\rm v}2}$ excluded. The
vacuum fluctuation, the exchange, and the Lamb-shift parts of the
two-loop
contribution are also shown.}
\label{thesis3.20}
\end{figure}
\begin{figure}
\caption{
The RHA and the two-loop contributions to the refitted energy density
with $F_{{\rm v}2}$ included.
The scalar, vector charge, and vector anomalous parts of the two-loop
contribution are also shown.}
\label{thesis3.21}
\end{figure}
\begin{figure}
\caption{
The vector charge form factor
$F_{{\rm v}1}$ at $M^{\scriptstyle\ast}=0.7M$. The solid
(dashed) line is obtained with polynomial (gaussian) interpolation.
The dotted
and dot-dashed lines are the corresponding vacuum form factors, which
are the same as those in Fig.~\protect\ref{fig:sv}(b).}
\label{vecmstar}
\end{figure}

\begin{figure}
\caption{
The perturbative two-loop energy with density-dependent vertices.
The couplings are the same as those in Fig.~\protect\ref{thesis3.14}.
 The dotted (dashed)
curves are obtained with $F_{{\rm v}2}$ excluded (included).
The solid curve is the RHA energy.}
\label{en_2lp_mst}
\end{figure}

\begin{figure}
\caption{
The baryon--scalar vertex that corresponds to
Eq.~(\protect\ref{eq:vtx}).
}
\label{fig:bsv}
\end{figure}
\vfill\eject
\mediumtext
\begin{table}
\caption{Interpolation parameter sets.}

\begin{tabular}{l l d d d d}
\multicolumn{2}{c}{\ \ \ \ } & \multicolumn{1}{c}{$\ \ \ \ \ \ \ \ a$} &
\multicolumn{1}{c}{$\ \ \ \ \ \ \ \ b$}
& \multicolumn{1}{c}{$\ \ \ \ \ \ \ \ c$}&
\multicolumn{1}{c}{$\ \ \ \ \ \ \ \ \ d$}   \\
\hline
   mixed   & $F_{\rm s}$ &0.747185  & 0.457293%
       & 2.60765 $ \times 10^{-1}$& $-$7.94998 $ \times 10^{-3}$   \\
               & $F_{{\rm v}1}$&0.792431 & 0.344247%
        & 2.12528 $ \times 10^{-1}$ &$-$4.95969 $ \times 10^{-3}$  \\
\hline
            & $F_{\rm s}$ &$-$1.30867 & 2.21488 & 9.87004 $ \times
               10^{-2}$ &$-$4.91063 $ \times 10^{-3}$\\
      poly. & $F_{{\rm v}1}$&$-$1.45088   & 2.38433%
& 6.92982 $ \times 10^{-2}$ &$-$2.75043 $\times 10^{-3}$\\
      &$F_{{\rm v}2}$&$-$2.94795 & 3.98240%
&  $-$3.35013 $ \times 10^{-2}$  &$-$9.41037 $\times 10^{-4}$\\
\hline
       & $F_{\rm s}$ &0.498343 &   0.541225%
& $-$4.14082 $ \times 10^{-2}$%
& 1.83978  $ \times 10^{-3}$\\
gauss.& $F_{{\rm v}1}$&0.348549   &0.720836%
&$-$7.39583 $ \times 10^{-2}$%
& 4.57337 $ \times 10^{-3}$\\
      &$F_{{\rm v}2}$ &$-$1.43842 &  2.83889%
& $-$4.19250 $\times 10^{-1}$%
& 1.87796 $ \times 10^{-2}$\\
\end{tabular}
\label{tab:abcd}
\end{table}

\begin{table}
\caption{Two-loop refitted parameter sets.}

\begin{tabular}{ c c c d d c c }
     set & interp. & $F_{{\rm v}2}$ & $g_{\rm s}^2$ &  $g _{\rm v}^2$
& $m_{\rm s} ({\,\rm MeV})$ &
$K^{-1}_{\rm v} ({\,\rm MeV})$  \\
\hline
    A  &  poly.  & No  &   63.6  &  130.8 & 458 &   600\\
    B  &  poly.  & Yes &  102.6  &  109.6 & 600 &   524\\
    C  &  gauss.  & No &  159.4  &  79.4  & 800 &   437\\
    D  &  gauss.  & Yes&  233.6  & 106.2  & 900 &   494 \\
\end{tabular}
\label{tab:paras}
\end{table}

\end{document}